\documentclass[sigconf]{acmart}

\usepackage{booktabs} 
\usepackage{subcaption}

\setcopyright{rightsretained}
  \usepackage{etoolbox}
 \usepackage{mathrsfs}
 \usepackage{mathtools}
\usepackage{balance}

\setcopyright{rightsretained}
\acmConference[SIGIR 2019 eCom]{ACM SIGIR Workshop on eCommerce}{July 2019}{Paris, France} 
\acmYear{2019}
\copyrightyear{2019}
\makeatletter
\renewcommand\@formatdoi[1]{\ignorespaces}
\makeatother
\acmISBN{}

\begin{document}
\title{Leverage Implicit Feedback for Context-aware Product Search}

\author{Keping Bi$^1$, Choon Hui Teo$^2$,  Yesh Dattatreya$^2$, Vijai Mohan$^2$, W. Bruce Croft$^1$}
\affiliation{%
	\institution{$^1$Center for Intelligent Information Retrieval, University of Massachusetts Amherst}
}
\email{{kbi, croft}@cs.umass.edu}
\affiliation{%
	\institution{$^2$Search Labs, Amazon}
}
\email{{choonhui, ydatta, vijaim}@amazon.com}


\begin{abstract}
Product search serves as an important entry point for online shopping. In contrast to web search, the retrieved results in product search not only need to be relevant but also should satisfy customers' preferences in order to elicit purchases. 
Previous work has shown the efficacy of purchase history in personalized product search \cite{ai2017learning}. However, customers with little or no purchase history do not benefit from personalized product search. Furthermore, preferences extracted from a customer's purchase history are usually long-term and may not always align with her short-term interests.
Hence, in this paper, we leverage clicks within a query session, as implicit feedback, to represent users' hidden intents, which further act as the basis for re-ranking subsequent result pages for the query. 
It has been studied extensively to model user preference with implicit feedback in recommendation tasks. However, there has been little research on modeling users' short-term interest in product search. We study whether short-term context could help promote users' ideal item in the following result pages for a query. Furthermore, we propose an end-to-end context-aware embedding model which can capture long-term and short-term context dependencies. 
Our experimental results on the datasets collected from the search log of a commercial product search engine show that short-term context leads to much better performance compared with long-term and no context. Our results also show that our proposed model is more effective than word-based context-aware models.


\end{abstract}

%
%

\keywords{Implicit Feedback, Product Search, Context-aware Search}

\maketitle

\section{Introduction}
\label{sec:introduction}
Online shopping has become an important part of people's daily life in recent years. In 2017, e-commerce represented 8.2\% of global retail sales (2,197 billion dollars); 
46.4\% of internet users shop online and nearly one-fourth of them do so at least once a week \cite{saleh2018}. 
Product search engines have become an important starting point for online shopping. A number of consumer surveys have shown that more online shoppers started searches on a retailer's online store (e.g., Amazon) rather than a generic web search engine (e.g., Google) \cite{Garcia2018}. 

In contrast to document retrieval, where relevance is a universal evaluation criterion, a product search system is evaluated based on user purchases that 
depend on both product relevance and customer preferences.
Previous research on product search \cite{duan2013probabilistic, duan2013supporting, van2016learning, wu2018turning, karmaker2017application} focused on product relevance. Several attempts \cite{parikh2011beyond, yu2014latent} were also made to improve customer satisfaction by diversifying search results. 
\citet{ai2017learning} introduced a personalized ranking model which takes the users' preferences learned from their historical reviews together with the queries as the basis for ranking. However, their work has several limitations. First, the personalized model cannot cope with the situations such as users that have not logged in during searching and thus can not be identified; users that logged in but do not have enough purchase history, and a single account being shared by several family members. In these cases, user purchase records are either not available or containing substantial noise. Second, given a specific purchase need expressed as a search query, long-term behaviors may not be as informative to indicate the user's preferences as short-term behaviors such as interactions with the retrieved results. These limitations of existing work on product search motivate us to model customers' preferences based on their interactions with search results, which do not require 
additional customers' information or their purchase history.

Customers' interactions with search results such as clicks
can be considered as implicit feedback based on their preferences. In information retrieval (IR), there are extensive studies on how to use users' feedback on the relevance of top retrieved documents to abstract a topic model and retrieve more relevant results \cite{rocchio1971relevance, lavrenko2017relevance, Zamani:2016:EQL:2970398.2970405}. These feedback techniques were shown to be very effective and can also be applied to use implicit feedback such as clicks. In contrast to document retrieval where a users' information need can usually be satisfied by a single click on a relevant result, we observe that, in product search, 
users tend to paginate to browse more products and make comparisons before they make final purchase decisions. 
In about 5\% to 15\% of search traffic, users browse and click results in the previous pages and purchase items in the later result pages.  
This provides us with the chance to collect user clicks more easily, based on which results shown in the next page can be tailored to meet the users' preferences. We reformulate product search as a dynamic ranking problem, where instead of one-shot ranking based on the query, the unseen products are re-ranked dynamically when users paginate to the next search result page (SERP) based on their implicit feedback collected from previous SERPs. 

Traditional relevance feedback (RF) methods, which extract word-based topic models from feedback documents as an expansion to the original queries, have potential word mismatch problems despite their effectiveness \cite{Zamani:2016:EQL:2970398.2970405,Rekabsaz:2016:GTM:2983323.2983833}. To tackle this problem, we propose an end-to-end context-aware embedding model that can incorporate both long-term and short-term context to predict purchased items. In this way, semantic match and the co-occurence relationship between clicked and purchased items are both captured in the embeddings. 
We show the effectiveness of incorporating short-term context against baselines using both no short-term context and word-based context. 

In this paper, we leverage implicit feedback as short-term context to provide users with more tailored search results. We first reformulate product search as a dynamic ranking problem, i.e., when users request next SERPs, the remaining unseen results will be re-ranked. We then introduce several context dependency assumptions for the task, and propose an end-to-end context-aware neural embedding model that can represent each assumption by changing the coefficients to combine long-term and short-term context. We further investigated the effect of several factors in the task: short-term context, long-term context, and neural embeddings. Our experimental results on the datasets collected from search logs of a commercial product search engine showed that incorporating short-term context leads to better performance compared with long-term context and no context, and embedding-based models perform better than word-based methods in the task under various settings. 

Our contributions can be summarized as follows: (1) we reformulate conventional one-shot ranking to dynamic ranking (i.e., multi-page search) based on user clicks in product search, which has not been studied before; (2) we introduce different context dependency assumptions and propose a simple yet effective end-to-end embedding model to capture different types of dependency; (3) we investigate different aspects in the dynamic ranking task on real search log data and confirmed the effectiveness of incorporating short-term context and neural embeddings. Our study on multi-page product search indicates that this is a promising direction and worth more attention. 
\section{Related Work}
\label{sec:related_work}
Next, we review three lines of research related to our work: product search, session-aware recommendation, and user feedback for information retrieval. 
\subsection{Product Search}
\label{subsec:product_search}
Product search has different characteristics compared with general web search; product information is usually more structured and the evaluation is usually based on purchases rather clicks. In 2006, \citet{jansen2006effectiveness} noted that the links retrieved by an e-commerce search engine are significantly better than those obtained from general search engines. Since the basic properties of products such as brands, categories and price are well-structured, considerable work has been done on searching products based on facets ~\cite{lim2010multi, vandic2013facet}. However, user queries are usually in natural language and hard to structure. To support keyword search, \citet{duan2013supporting, duan2013probabilistic} extended the Query Likelihood method \cite{ponte1998language} by considering the query generated from a mixture of the language model of background corpus and the language model of the products conditioned on their specifications. The ranking function constructed in this approach utilizes exact word matching information whereas vocabulary mismatch between free-form user queries and product descriptions or reviews from other users can still be an issue. \citet{van2016learning} noticed this problem and introduced a latent vector space model which matches queries and products in the semantic space. The latent vectors of products and words are learned in an unsupervised way, where vectors of n-grams in the description and reviews of the product are used to predict the product. Later, \citet{ai2017learning} built a hierarchical embedding model
in which, learned representations of users, queries, and products are used to predict product purchases and associated reviews. 


Other aspects of product search such as popularity, visual preference and diversity have also been studied. \citet{li2011towards} investigated product retrieval from an economic perspective. \citet{long2012enhancing} predicted sales volume of items based on their transaction history and incorporate this complementary signal with relevance for product ranking. The effectiveness of images for product search was also investigated \cite{di2014relevance,guo2018multi}. To satisfy different users' intents behind the same query, efforts on improving result diversity in product retrieval have also been made \cite{parikh2011beyond, yu2014latent}. 

In terms of labels for training, there are studies on using clicks as an implicit feedback signal. \citet{wu2018turning} jointly modeled clicks and purchases in a learning-to-rank framework in order to optimize the gross merchandise volume. To model clicks, they consider click-through rate of an item for a given query in a set of search sessions as the signal for training. \citet{karmaker2017application} compared the different effects of exploiting click-rate, add-to-cart ratios, order rates as labels. They experimented on multiple representative learning to rank models in product search with various settings. Our work also uses clicks as implicit feedback signals, but instead of aggregating all the clicks under the same query to get click-through rate, we consider the clicks associated with each query as an indicator of the user's short-term preference behind that query. 

Most previous work treat product search as a one-shot ranking problem, where given a query, static results are shown to users regardless of their interaction with the result lists. In a different approach, \citet{hu2018reinforcement} formulate the user behaviors during searching products as a Markov decision process (MDP) and use reinforcement learning to optimize the accumulative gain (expected price) of user purchases. They define the states in the MDP to be a non-terminal state, from where users continue to browse, and two terminal states, i.e. purchases happen (conversion events) or users abandon the results (abandon events). Their method is essentially online learning and refines the ranking model with large-scale users' behavior data. Although we work on a similar scenario where the results shown in next page can be revised, they gradually refine an overall ranker that affects all the queries while our model revises results for each individual query based on the estimation of the user preference under the query. Another difference is that they only consider purchases as a deferred signal for training and do not use any clicks in the process. In contrast, we treat clicks as an indicator of user preferences and refine ranking conditioned on the preferences. 
\subsection{Session-aware Recommendation}
\label{subsec:session_rec}
In session-aware recommendation, a user's interactions with the previously seen items in the session are used for recommending the next item. 
Considerable research on session-aware recommendation has been done in the application domains such as news, music, movies and products. Many these works are based on matrix factorization \cite{rendle2011fast,hidasi2016general, jawaheer2014modeling}.  More recently, session-aware recommendation approaches based on neural networks have shown superior performance. \citet{hidasi2015session} model the clickstream in a session with Gated Recurrent Unit (GRU) and predict the next item to recommend in the session.  \citet{twardowski2016modelling} also used Recurrent Neural Networks (RNN) but used attributes for item encoding and recommended only on unseen items. 
\citet{quadrana2017personalizing} proposed a hierarchical RNN model, which consists of a session-level GRU to model users' activities within sessions and a user-level GRU to model the evolution of the user across sessions. The updated user representation will affect the session-level GRU to make personalized recommendations.
\citet{wu2017session} proposed a two-step ranking method to recommend item lists based on user clicks and views in the session. They treat item ranking as a classification problem and learn the session representation in the first step. With the session representation as context, items are reranked with a list-wise loss proposed in ListNet in the second step. \citet{li2017neural} adopted the attention mechanism in the RNN encoding process to identify the user's main purpose in the current session. \citet{quadrana2018sequence} reviewed extensive previous work on sequence-aware recommendation and categorized the existing methods in terms of different tasks, goals, and types of context adaption. 

The goal of a recommendation system is typically to help users explore items that they may be interested in when they do not have clear purchase needs. On the contrary, a search engine aims to help users find only items that are most relevant to their intent specified in search queries. Relevance plays totally different roles in the two tasks. 
In addition, the evaluation metrics in recommendation are usually based on clicks \cite{hidasi2015session, quadrana2017personalizing,twardowski2016modelling,wu2017session,li2017neural}, whereas product search is evaluated with purchases under a query. 


\subsection{User Feedback for Information Retrieval}
\label{subsec:feedback_IR}
There are studies on two types of user feedback in information retrieval, implicit feedback which usually considers click-through data as the indicator of document relevance and explicit feedback where users are asked to give the relevance judgments of a batch of documents.  \citet{joachims2017accurately} found that click-through data as implicit feedback is informative but biased and the relative preferences derived from clicks are accurate on average. To separate click bias from relevance signals, \citet{craswell2008experimental} designed a Cascade Model by assuming that users examine search results from top to bottom; \citet{dupret2008user} proposed a User Browsing Model where results can be skipped according to their examination probability estimated from their positions and last clicks; \citet{chapelle2009dynamic} constructed a Dynamic Bayesian Network model which incorporate a variable to indicate whether a user is satisfied by a click and leaves the result page. \citet{yue2009interactively} defined a dueling bandit problem where reliable relevance signals are collected from users' clicks on interleaved results to optimize the ranking function. Learning an unbiased model directly from biased click-through data has also been studied by incorporating inverse propensity weighting and estimating the propensity \cite{wang2018position, joachims2017unbiased, ai2018unbiased}. In this work, we model the user preference behind a search query with her clicks and refine the following results shown to this user. 

Explicit feedback is also referred to as true relevance feedback (RF) in information retrieval and has been extensively studied. Users are asked to assess the relevance of a batch of documents based on which the retrieval model is refined to find more relevant results. Rocchio \cite{rocchio1971relevance} is generally credited as the first relevance feedback method, which is based on the vector space model \cite{salton1975vector}. After the language model approach for IR has been proposed \cite{ponte1998language},  the relevance model version 3 (RM3) \cite{lavrenko2017relevance} became one of the state-of-art pseudo RF methods that is also effective for relevance feedback. \citet{Zamani:2016:EQL:2970398.2970405} incorporate the semantic match between unsupervised trained word embeddings into the language model framework and introduced an embedding-based relevance model (ERM). Although these RF methods can also be applied in our task, we propose an end-to-end neural model for relevance feedback in the context of product search. 

\section{Context-aware Product Search}
\label{sec:context_ps}
We reformulate product search as a dynamic re-ranking task where short-term context represented by the clicks in the previous SERPs is considered for re-ranking subsequent result pages. Users' global interests can also be incorporated for re-ranking as long-term context. We first introduce our problem formulation and different assumptions of context dependency models. Then we propose a context-aware embedding model for the task and show how to optimize the model.

\subsection{Problem Formulation}
\label{subsec:prob_form}
A query session\footnote{We refer to the series of user behaviors associated with a query as  a query session, i.e, a user issues a query, clicks results, paginates, purchases items and finally ends searching with the query. } is initiated when a user $u$ issues a query $q$ to the search engine. The search results returned by the search engine are typically grouped into pages with similar number of items.
Let $R_t$ be the set of items on the $t$-th search result page ranked by an initial ranker and denote by $R_{1:t}$ the union of $R_1, \cdots, R_t$. For practical purposes, we let the re-ranking candidate set $D_{t+1}$ for page $t+1$ be $R_{1:t+k}  \diagdown V_{1:t}$ where $k \geq 1$ and $V_{1:t}$ is the set of re-ranked items viewed by the user in the first $t$ pages. Given user $u$, query $q$, and the set of clicked items in the first $t$ pages $C_{1:t}$ as context, the objective is to rank all, if any, purchased items $B_{t+1}$ in $D_{t+1}$ at the top of the next result page. 

\subsection{Context Dependency Models}
\label{subsec:depend_context}
\begin{figure}
	\includegraphics[width=0.48\textwidth]{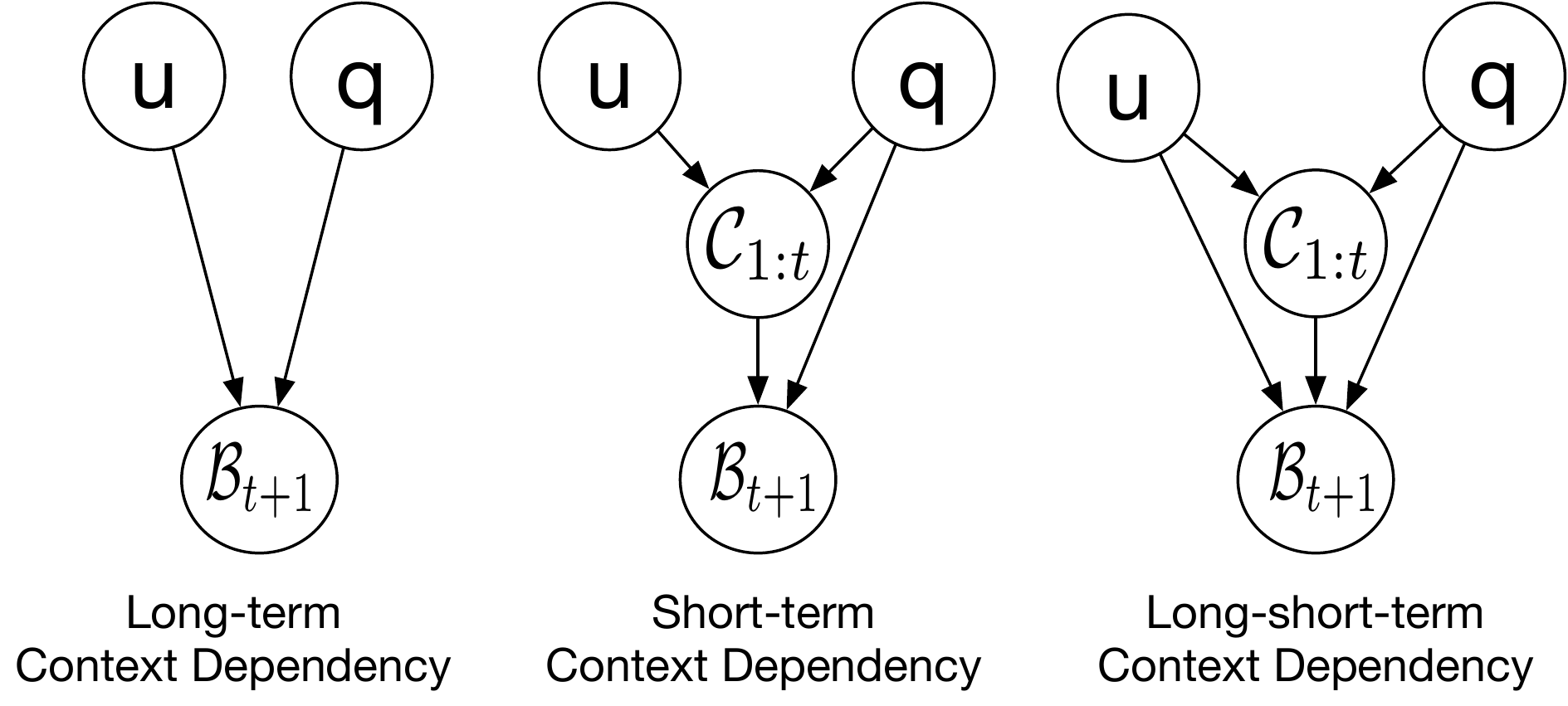} %
	\caption{Different assumptions to model different factors as context for purchase prediction.}
	\label{fig:depend}
\end{figure}

There are three types of context dependencies that one can use to model the likelihood of a user purchasing a product in her query session, namely, long-term context, short-term context, and long-short-term context. Figure \ref{fig:depend} shows the graphical models for these context dependencies, where $u$ denotes the latent variable of a user's long-term interest that stays the same across all the search sessions, and clicks in the first $t$ result pages, i.e., $C_{1:t}$, represents the user's short-term preference. Purchased items on and after page $t+1$, i.e., $\mathcal{B}_{t+1}$, depends on query $q$ and different types of context under different dependency assumptions. 

\textbf{Long-term Context Dependency.} 
In this assumption, only users' long-term preferences, usually represented by their historical queries and the corresponding purchased items, are used to predict the purchases in their current query sessions. An unshown item $i$ is ranked according to its probability of being purchased given $u$ and $q$, namely $p(i \in \mathcal{B}_{t+1} | u, q)$. The advantage of such models is that personalization of search results (as proposed in \citet{ai2017learning}) can be conducted from the very beginning of a query session when there is no feedback information available. However, this model needs user identity and purchase history, which are not always available. In addition, the long-term context may not be informative to predict a user's final purchases since her current search intent may be totally different from any of her previous searches and purchases. 


\textbf{Short-term Context Dependency.}
The shortcomings of long-term context can be addressed by focusing on just the short-term context, i.e., the user's actions such as clicks performed within the current query session. This dependency model assumes that given the observed clicks in the first $t$ pages, the items purchased in the subsequent result pages are conditionally independent of the user, shown in Figure \ref{fig:depend}. 
An unseen item $i$ in the query session is re-ranked based on its purchase probability conditioning on $\mathcal{C}_{1:t}$ and $q$, i.e., $p(i \in \mathcal{B}_{t+1} | \mathcal{C}_{1:t}, q)$. 
In this way, users' short-term preferences are captured and their identity and purchase records are not needed. Users with little or no purchase history and who have not logged in can benefit directly under such a ranking scheme.

\textbf{Long-short-term Context Dependency.}
The third dependency assumption is that purchases in the subsequent result pages depend on both short-term context, e.g., previous clicks in the current query session, and long-term context, such as historical queries and purchases of the user indicated by $u$. An unseen item $i$ after page $t$ is scored according to $p(i \in \mathcal{B}_{t+1} | \mathcal{C}_{1:t}, q, u)$. This setting considers more information but it also has the drawback of requiring users identity and purchase history. 

We will introduce how to model the three dependency assumptions in a same framework in Section \ref{subsec:embed_ca_model}. In this paper, we focus on the case of non-personalized short-term context and include the other two types of context for comparison.

\subsection{Context-aware Embedding Model}
\label{subsec:embed_ca_model}
\begin{figure}
	\includegraphics[width= 0.49 \textwidth]{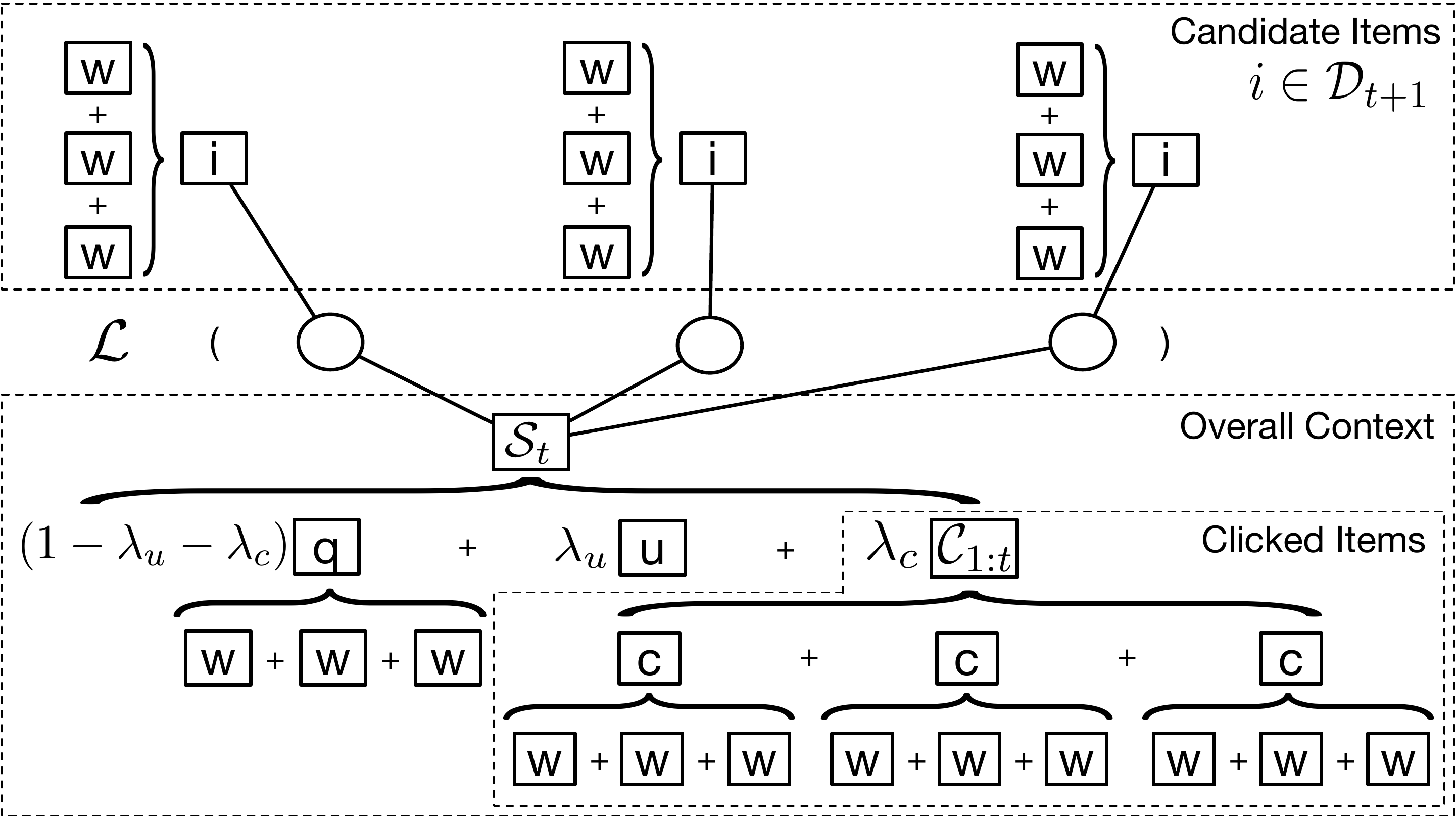} %
	\caption{The structure of our context-aware embedding model (CEM). $w$ represents words in queries or product titles; $\mathcal{C}_{1:t}$ denotes the click item set in the first $t$ SERPs, which consist of item $c$; $\mathcal{S}_t$ is the overall context of the first $t$ SEPRs, a combination of query $q$, user $u$ and clicks $\mathcal{C}_{1:t}$; $i$ is an item in the candidate set $\mathcal{D}_{t+1}$ for re-ranking from page $t+1$. }
	\label{fig:model}
\end{figure}

We designed a context-aware framework where models under different dependency assumptions can be trained by varying the corresponding coefficients, shown in Figure \ref{fig:model}. 
To incorporate semantic meanings and avoid the word mismatch between queries and items, we embed queries, items and users into latent semantic space.
Our context-aware embedding model is referred to as CEM.
We assume users' preferences are reflected by their implicit feedback, i.e. their clicks associated with the query. Similar to relevance feedback approaches \cite{lavrenko2017relevance, rocchio1971relevance} that extract a topic model from assessed relevant documents, our model should capture user preferences from their clicked items which are implicit positive signals. Components of CEM will be introduced next. 

\textbf{Item Embeddings.}
We use product titles to represent products since merchants tend to put the most informative, representative text such as the brand, name, size, color, material and even target customers in product titles. In this way, items do not have unique embeddings according to their identifiers and items with the same titles are considered the same. Although this may not be accurate all the time, word representations can be generalized to new items, and we do not need to cope with the cold-start problem.
We use the average of title word embeddings of a product as its own embedding, i.e.,
\begin{equation}
\label{eq:item_embed}
\mathcal{E}(i)= \frac{\sum_{w \in i} \mathcal{E}(w)}{|i|}
\end{equation}
where $i$ is the item, and $|i|$ is the title length of item $i$.
We also evaluated other more complex product title encoding approaches such as non-linear projection of average word embeddings and recurrent neural network on title word sequence, but they did not show superior performance over the simpler one that we use here.

\textbf{User Embeddings.}
A lookup table for user embeddings is created and used for training, where each user has a unique representation. This vector is shared across search sessions and updated by the gradient learned from previous user transactions. In this way, the long-term interest of the user is captured and we use the user embeddings as long-term context in our models.

\textbf{Query Embeddings.}
Similar to item embeddings, we use the simple average embedding of query words as the representation, which also shows the best performance compared to the non-linear projection and recurrent neural network methods we have tried. The embedding of the query is 
\begin{equation}
\mathcal{E}(q)= \frac{\sum_{w \in q} \mathcal{E}(w)}{|q|}
\end{equation}
where $|q|$ is the length of query $q$.

\textbf{Short-term Context Embeddings. }
We use the set of clicked items to represent user preference behind the query, which we refer to as $\mathcal{E}(\mathcal{C}_{1:t})$. For sessions associated with a different query $q$ or page number $t$, the clicked items contained in $\mathcal{C}_{1:t}$ may differ. 
We assume the sequence of clicked items does not matter when modeling short-term user preference, i.e., the same set of clicked items should imply the same user preference regardless of the order of them being clicked. There are two reasons for this assumption. One is that the user's purchase need is fixed for a query she issued and is not affected by the order of clicks. The other is that the order of user clicks is usually based on the rank of retrieved products from top to bottom as the user examines each result, which is not affected by user preference in the non-personalized search results. 
So we represent the set as the centroid of each clicked item in the latent semantic space, where the order of clicks does not make a difference. 
A simple yet effective way is to consider equal weights of all the items in $\mathcal{C}_{1:t}$ so that the centroid is simply averaged item embeddings:
\begin{equation}
\mathcal{E}(\mathcal{C}_{1:t}) = \frac{\sum_{i \in \mathcal{C}_{1:t}}\mathcal{E}(i)}{|\mathcal{C}_{1:t}|}
\end{equation}
where $|\mathcal{C}_{1:t}|$ is the number of clicked items in set $\mathcal{C}_{1:t}$. 

We also tried an attention mechanism to weight each clicked item according to the query and represent the user preference with a weighted combination of clicked items. However, this method is not better than combining clicks with equal weights in our experiments. So we only show simple methods. 

\textbf{Overall Context Embeddings.}
We use a convex combination of user, query, and click embeddings as the representation of overall context $\mathcal{E}(\mathcal{S}_t)$. i.e.
\begin{equation}
\label{eq:context}
\begin{aligned}
\mathcal{E}(\mathcal{S}_t) &= (1-\lambda_u-\lambda_c) \mathcal{E}(q) + \lambda_u \mathcal{E}(u) + \lambda_c \mathcal{E}(\mathcal{C}_{1:t}) \\
& 0 \leq \lambda_u \leq 1, 0 \leq \lambda_c \leq 1, \lambda_u + \lambda_c \leq 1 \\
\end{aligned}
\end{equation}
This overall context is then treated as the basis for predicting purchased items in $\mathcal{B}_{t+1}$. 
When $\lambda_c = 0$, $\mathcal{C}_{1:t}$ is ignored in the prediction and $\mathcal{S}_t$ corresponds to the long-term context shown in Figure \ref{fig:depend}. When $\lambda_u=0$, user $u$ does not have impact on the final purchase given $\mathcal{C}_{1:t}$. This aligns with the short-term context assumption in Figure \ref{fig:depend}. When $\lambda_u > 0, \lambda_c > 0, \lambda_u + \lambda_c \leq 1$, both long-term and short-term context are considered and this matches the type of long-short-term context in Figure \ref{fig:depend}. 
So by varying the values of $\lambda_u$ and $\lambda_c$, we can use Equation \ref{eq:context} to model different types of context dependency and do comparisons.

\textbf{Attention Allocation Model for Items. } 
With the overall context collected from the first $t$ pages, we further construct an attentive model to re-rank the products in the candidate set $\mathcal{D}_{t+1}$. This re-ranking process can be considered as an attention allocation problem. Given the context that indicates the user's preference and a set of candidate items that have not been shown to the users yet, the item which attracts more user attention will have higher probability to be purchased. The attention weights then act as the basis for re-ranking. Predicting the probability of each candidate item being purchased can be considered as attention allocation for the items. This idea is also similar to the listwise context model proposed by \citet{ai2018learning}. They extracted the topic model from top-ranked documents with recurrent neural networks and used it as a local context to re-rank the top documents with their attention weights. The attention weights can be computed as:
\begin{equation}
\label{eq:attn_score}
\begin{aligned}
score(i | q,u,\mathcal{C}_{1:t}) = \frac{\exp(\mathcal{E}(\mathcal{S}_t) \cdot \mathcal{E}(i))}
{\sum_{i' \in \mathcal{D}_{t+1}}\exp( \mathcal{E}(\mathcal{S}_t) \cdot \mathcal{E}(i'))} \\
\end{aligned}
\end{equation}
where $\mathcal{E}(\mathcal{S}_t)$ is computed according to Equation \ref{eq:context}. 
This model can also be interpreted as a generative model for an item in the candidate set $\mathcal{D}_{t+1}$ given the context $\mathcal{S}_{t}$. 
In this case, the probability of an item in the candidate set $\mathcal{D}_{t+1}$ being generated from the context $\mathcal{S}_{t}$ is computed with a softmax function that take the dot product score between the embedding of an item and the context as inputs, i.e,
\begin{equation}
\label{eq:condition_prob}
p(i|\mathcal{C}_{1:t}, u, q) = score(i | q,u,\mathcal{C}_{1:t}) 
\end{equation}
We need to train the model and learn appropriate embeddings of context and items so that the probability of purchased items in $\mathcal{D}_{t+1}$, namely $\mathcal{B}_{t+1}$, should be larger than the other candidate items, i.e. $\mathcal{D}_{t+1} \diagdown \mathcal{B}_{t+1}$. Also, the conditional probability in Equation \ref{eq:condition_prob} can be used to compute the likelihood of the observed instance of $\mathcal{C}_{1:t}, u, q, \mathcal{B}_{t+1}$. 

\subsection{Model Optimization}
\label{subsec:model_opt}
The embeddings of queries, users, items are learned by maximizing the likelihood of observing $\mathcal{B}_{t+1}$ given the condition of $\mathcal{C}_{1:t}, u, q$, i.e., after user $u$ issued query $q$, she clicked the items in the first $t$ SERPs ($\mathcal{C}_{1:t}$), then models are learned by maximizing the likelihood for her to finally purchased items in $\mathcal{B}_{t+1}$ which are shown in and after page $t+1$. There are many possible values of $t$ even for a same user $u$ if she purchases multiple products on different result pages under query $q$. These are considered as different data entries. Then the log likelihood of observing purchases in $\mathcal{B}_{t+1}$ conditioning on $\mathcal{C}_{1:t}, u, q$ in our model can be computed as
\begin{equation}
\label{eq:likelihood}
\begin{aligned}
\mathcal{L}(\mathcal{B}_{t+1} | \mathcal{C}_{1:t}, u, q) 
&= \log p(\mathcal{B}_{t+1} | \mathcal{C}_{1:t}, u, q)  \\
& \propto \log \prod_{i \in \mathcal{B}_{t+1}} p(i | \mathcal{C}_{1:t}, u, q) \\
& \propto \sum_{i \in \mathcal{B}_{t+1}} \log p(i | \mathcal{C}_{1:t}, u, q) \\
\end{aligned}
\end{equation}
The second step can be inferred if we consider whether an item will be purchased is independent of another item given the context. 

According to Equation \ref{eq:attn_score}, \ref{eq:condition_prob} and \ref{eq:likelihood}, we can optimize the conditional log-likelihood directly. A common problem for the softmax calculation is that the denominator usually involves a large number of values and is impractical to compute. However, this is not a problem in our model since we limit the candidate set $\mathcal{D}_{t+1}$ to only some top-ranked items retrieved by the initial ranker so that the computation cost is small. 

Similar to previous studies \cite{van2016learning, ai2017learning}, we apply L2 regularization on the embeddings of words and users to avoid overfitting. The final optimization goal can be written as 
\begin{equation}
\begin{split}
\mathcal{L'} = &\sum_{u, q, t} \mathcal{L}(\mathcal{B}_{t+1}| \mathcal{C}_{1:t}, u, q) 
+ \gamma (\sum_{w} {\mathcal{E}(w)}^2 + \sum_{u} {\mathcal{E}(u)}^2) \\
= &\sum_{u, q, t}  \sum_{i \in \mathcal{B}_{t+1}} \log 
\frac{\exp(\mathcal{E}(\mathcal{S}_t) \cdot \mathcal{E}(i))}
{\sum_{i' \in \mathcal{D}_{t+1}}\exp( \mathcal{E}(\mathcal{S}_t) \cdot \mathcal{E}(i'))} \\
& \; \; \; + \gamma \big(\sum_{w} {\mathcal{E}(w)}^2 + \sum_{u} {\mathcal{E}(u)}^2 \big) \\ 
\end{split}
\end{equation}
where $\gamma$ is the hyper-parameter to control the strength of L2 regularization. 
The function accumulates entries of all the possible user $u$, query $q$, and the valid page number $t$ for pagination which has clicks in and before page $t$ and purchases after that page. All possible words and users are taken into account in the regularization. When we do not incorporate long-term context, the corresponding parts of $u$ are omitted.  

The loss function actually captures the loss of a list and this list-wise loss is similar to AttentionRank proposed by \citet{ai2018learning}.  Because of the softmax function, optimizing the probabilities of relevant instances in $\mathcal{B}_{t+1}$ simultaneously minimizes the probabilities of the rest non-relevant instances. This loss shows superiority over other list-wise loss such as ListMLE \cite{xia2008listwise} and SoftRank \cite{taylor2008softrank}, which is another reason we adopt this loss.

\section{Experimental Setup}
\label{sec:exp_setup}
In this section, we introduce our experimental settings of context-aware product search. We first describe how we construct the datasets for experiments. Then we describe the baseline methods and evaluation methodology for comparing different methods. We also introduce the training settings for our model.
\subsection{Datasets}
\label{subsec:datasets}

\begin{table}
	\caption{Statistics of our collected datasets }
	\centering
	\label{tab:stats}    
	\small
	\begin{tabular}{l  r  r  r }
		\hline
		& Toys & Garden & Cell Phones \\
		& \& Games & \& Outdoor & \& Accessories \\
		\hline
		Product title length & 13.14$\pm$6.46 & 16.39$\pm$7.38 & 22.02$\pm$7.34 \\
		Vocabulary size & 381,620 & 1,054,980 & 194,022 \\
		\hline
		\multicolumn{4}{l}{Query Session Splits}  \\
		Train & 91.21\%  &	87.36\% & 86.57\%  \\
		Validation & 2.61\% &	3.66\% & 4.20\% \\
		Test & 6.18\% &	8.98\% & 9.23\%  \\
		\hline
	\end{tabular}
\end{table}

We randomly sampled three category-specific datasets, namely, ``Toys \& Games'', ``Garden \& Outdoor'', and  ``Cell Phones \& Accessories'', from the logs of a commercial product search engine spanning ten months between years 2017 and 2018. 
We keep only the query sessions with at least one clicked item on any page before the pages with purchased items. These sessions are difficult for the production model since it could not rank the ``right'' items on the top so that users purchased items in the second or later result pages.
Our datasets include up to a few million query sessions containing several hundred thousand unique queries.
When there are multiple purchases in a query session across different result pages, purchases until page $t$ are only considered as clicks and used together with other clicks to predict purchases on and after page $t+1$. 
Statistics of our datasets are shown in Table \ref{tab:stats}.

\subsection{Evaluation Methodology}
\label{subsec:eval_method}
We divided each dataset into training, validation, and test sets by the date of the query sessions. The sessions occurred in the first 34 weeks are used for training, the following 2 weeks for validation and the last 4 weeks for testing. 
Models were trained with data in the training set; hyper-parameters were tuned according to the model performance on the validation set, and evaluation results on the test set were reported for comparison. 

Since the datasets are static, it is impossible to evaluate the models in a truly interactive setting where each subsequent page is re-ranked based on the observed clicks on the current and previous pages. Nonetheless,
we can still evaluate the performance of one-shot re-ranking from page $t+1$ given the context collected from the first $t$ pages. 
In our experiments, we compare different methods for re-ranking from page 2 and page 3 since earlier re-ranking can influence results at higher positions which have bigger larger impact on the ranking performance.
As in relevance feedback experiments \cite{lv2009adaptive, rocchio1971relevance}, our evaluation is also based on residual ranking, where the first $t$ result pages are discarded and re-ranking of the unseen items are evaluated. We use the residual ranking evaluation paradigm because the results before re-ranking are retrieved by the same initial ranker and identical for all the re-ranking methods. 

Similar to other ranking tasks, we use mean average precision ($MAP$) at cutoff 100, mean reciprocal rank ($MRR$) and normalized discounted cumulative gain ($NDCG$) as ranking metrics. 
$MAP$ measure the overall performance of a ranker in terms of both precision and recall, which indicates the ability to retrieve more purchased items in next 100 results and ranking them to higher positions. 
$MRR$ is the average inverse rank for the first purchase in the retrieved items. It indicates the expected number of products users need to browse before finding the ones they are satisfied with. 
$NDCG$ is a common metric for multiple-label document ranking. Although in our context-aware product search, items only have binary labels indicating whether they were purchased given the context, $NDCG$ still shows how good a rank list is with emphasis on results at top positions compared with the ideal rank list. We use $NDCG@10$ in our experiments. 
\subsection{Baselines}
\label{subsec:baseline}
We compare our short-term context-aware embedding model (SCEM) with four groups of baseline, retrieval model without using context, long-term, short-term and long-short-term context-aware models. 

\textbf{Production Model (PROD).}
PROD is essentially a gradient boosted decision tree based model. Comparing with this model indicates the potential gain of our model if deployed online. Note that PROD performs worse on our datasets than on the entire search traffic since we extracted query sessions where the purchased items are in the second or later result pages. 

\textbf{Random (RAND).}
By randomly shuffling the results in the candidate set which consists of the top unseen retrieved items by the production model, we get the performance of a random re-ranking strategy. This performance should be the lower bound of any reasonable model. 

\textbf{Popularity (POP).}
In this method, the products in the candidate set are ranked according to how many times they were purchased in the training set. Popularity is an important factor for product search \cite{long2012enhancing} besides relevance. 

\textbf{Query Likelihood Model (QL).}
The query likelihood model (QL) \cite{ponte1998language} is a language model approach for information retrieval. 
It shows the performance of re-ranking without implicit feedback and is only based on the bag-of-words representation. The smoothing parameter $\mu$ in QL was tuned from $\{10, 30, 50, 100, 300, 500\}$. 

\textbf{Query Embedding based Model (QEM).}
This model scores an item by the generative probability of the item given the embedding of a query. When $\lambda_u = 0, \lambda_c = 0$, $CEM$ is exactly $QEM$. 

\textbf{Long-term Context-aware Relevance Model (LCRM3).}
Relevance Model Version 3 (RM3) \cite{lavrenko2017relevance} is an effective method for both pseudo and true relevance feedback. It extracts a bag-of-words language model from a set of feedback documents, expands the original query with the most important words from the language model, and retrieve results again with the expanded query. To capture the long-term interest of a user, we use RM3 to extract significant words from titles of the user's historical purchased products and refine the retrieval results for the user in the test set with the expanded query. 
The weight of the initial query was tuned from $\{0, 0.2, \cdots, 1.0\}$ and the expansion term count was tuned from $\{10, 20, \cdots, 50\}$. The effect of query weight is shown in Section \ref{subsec:query_effect}.  

\textbf{Long-term Context-aware Embedding Model (LCEM).}
When $\lambda_c = 0, 0 < \lambda_u \leq 1$, $CEM$ becomes $LCEM$ by considering long-term context indicated by universal user representations. 

\textbf{Short-term Context-aware Relevance Model (SCRM3).}
We also use RM3 to extract the user preference behind a query from the clicked items in the previous SERPs as short-term context and refine the next SERP. This method uses the same information as our short-term context-aware embedding model, but it represents user preference with a bag-of-words model and only -consider word exact match between a candidate item and the user preference model. The query weight and expansion term count were tuned in the same range as LC-RM3 and the influence of initial query weight can be found in Section \ref{subsec:query_effect}.
\footnote{We also implemented the embedding-based relevance model (ERM) \cite{Zamani:2016:EQL:2970398.2970405}, which is an extension of RM3 by taking semantic similarities between word embeddings into account, as a context-aware baseline. But it does not perform better than RM3 across different settings. So we did not include it.  }

\textbf{Long-short-term Context-aware Embedding Model (LSCEM).}
When $\lambda_u > 0, \lambda_c > 0, 0 < \lambda_u + \lambda_c \leq 1$, both long-term context represented by $u$ and short-term context indicated by  $\mathcal{C}_t$ are taken into account in $CEM$. 

PROD, RAND, POP, QL, and QEM are retrieval models that rank items based on queries and do not rely on context or user information.
These models can be used as the initial ranker for any queries.
The second type of rankers consider users' long-term interests together with queries, such as LCEM and LCRM3. These methods utilize users' historical purchases but can only be applied to users who appear in the training set. 
The third type is feedback models which take users' clicks in the query session as short-term context and this category includes SCRM3 and our SCEM. In this approach, user identities are not needed. However, they can only be applied to search sessions where users click on results and only items from the second result page or later can be refined with the clicks. 
The fourth category considers both long and short-term context, e.g., LSCEM. 
The second, third and fourth groups of baseline correspond to the dependency assumptions shown in the first, second and third sub-figure in Figure \ref{fig:depend} respectively. 

\subsection{Model Training}
\label{subsec:model_train}
Query sessions with multiple purchases on different pages are split into sub-sessions, one for each page with a purchase. 
When there are more than three sub-sessions for a given session, we randomly select three in each training epoch. We do so to avoid skewing the dataset with sessions with many purchases. Likewise, we randomly select five clicked items for constructing short-term context if there are more than five clicked items in a query session.

We implemented our models with
Tensorflow. The models were trained for 20 epochs with the batch size set to 256. 
Adam \cite{kingma2014adam} was used as the optimizer and the global norm of parameter gradients was clipped at 5 to avoid unstable gradient updates. 
After each epoch, the model was evaluated on the validation set and the model with the best performance on the validation set was selected to be evaluated on the test set. 
The initial learning rate was selected from $\{0.01, 0.005, 0.001, 0.0005, 0.0001\}$. L2 regularization strength $\gamma$ was tuned from 0.0 to 0.005. $\lambda_q, \lambda_u$ in Equation \ref{eq:context} were tuned from $\{0, 0.2, \cdots, 0.8, 1.0\}$ ($\lambda_q + \lambda_u \leq 1$) to represent various dependency assumptions mentioned in Section \ref{subsec:depend_context}, and the embedding size were scanned from $\{50, 100, \cdots, 300\}$. The effect of $\lambda_q$, $\lambda_u$ and embedding size are shown in Section \ref{sec:results}.


\section{Results and Discussion}
\label{sec:results}

In this section, we show the performance of the four types of models mentioned in Section \ref{subsec:baseline}. 
First, we compare the overall retrieval performance of various types of models in Section \ref{subsec:overall_perf}. 
Then we further study the effect of queries, long-term context and embedding size on each model in the following subsections.

\begin{table*}
	\caption{Comparison of baselines and our short-term context embedding model (SCEM) on re-ranking when users paginate to the 2nd and 3rd page. The number is the relative improvement of each method compared with the production model (PROD)\protect\footnotemark. `$^-$' indicates significant worse of each baseline compared with SCEM in student t-test with $p \leq 0.001$. Note that difference larger than 3\% is approximately significant. The best performance in each column is marked in bold. }
	\label{tab:overallperf}
	\scalebox{0.9}{    
	\begin{tabular}{  p{1.5cm} || r | r | r || r | r | r || r | r | r   }
		\hline
		\multicolumn{10}{c}{Performance of Re-ranking from the 2nd Page} \\
		\hline
		& \multicolumn{3}{c||}{Toys \& Games} & \multicolumn{3}{c||}{Garden \& Outdoor} & \multicolumn{3}{c}{Cell Phones \& Accessories} \\
		\hline
		Model & $MAP$ & $MRR$ & $NDCG@10$ & $MAP$ & $MRR$ & $NDCG@10$ & $MAP$ & $MRR$ & $NDCG@10$ \\
		\hline
		\hline
		PROD & 0.00\%$^{-}$ & 0.00\%$^{-}$ & 0.00\%$^{-}$ & 0.00\%$^{-}$ & 0.00\%$^{-}$ & 0.00\%$^{-}$ & 0.00\%$^{-}$ & 0.00\%$^{-}$ & 0.00\%$^{-}$ \\
		\hline
		RAND & -25.70\%$^{-}$ & -26.83\%$^{-}$ & -29.23\%$^{-}$ & -23.40\%$^{-}$ & -24.16\%$^{-}$ & -25.73\%$^{-}$ & -20.15\%$^{-}$ & -20.93\%$^{-}$ & -22.73\%$^{-}$ \\
		\hline
		POP & -15.82\%$^{-}$ & -15.90\%$^{-}$ & -17.87\%$^{-}$ & -9.38\%$^{-}$ & -9.51\%$^{-}$ & -9.55\%$^{-}$ & -8.54\%$^{-}$ & -8.25\%$^{-}$ & -11.12\%$^{-}$ \\
		\hline
		QL & -25.78\%$^{-}$ & -27.80\%$^{-}$ & -29.73\%$^{-}$ & -19.62\%$^{-}$ & -20.78\%$^{-}$ & -21.63\%$^{-}$ & -16.14\%$^{-}$ & -16.77\%$^{-}$ & -18.00\%$^{-}$ \\
		\hline
		QEM & -2.57\%$^{-}$ & -3.10\%$^{-}$ & -3.85\%$^{-}$ & +0.65\%$^{-}$ & -0.34\%$^{-}$ & +1.06\%$^{-}$ & +9.96\%$^{-}$ & +9.73\%$^{-}$ & +10.58\%$^{-}$ \\
		\hline
		\hline
		LCRM3 & -24.82\%$^{-}$ & -25.92\%$^{-}$ & -28.60\%$^{-}$ & -19.33\%$^{-}$ & -20.45\%$^{-}$ & -21.28\%$^{-}$ & -15.44\%$^{-}$ & -16.07\%$^{-}$ & -17.38\%$^{-}$ \\
		\hline
		LCEM & -2.57\%$^{-}$ & -3.10\%$^{-}$ & -3.85\%$^{-}$ & +0.65\%$^{-}$ & -0.34\%$^{-}$ & +1.06\%$^{-}$ & +9.96\%$^{-}$ & +9.73\%$^{-}$ & +10.58\%$^{-}$ \\
		\hline
		\hline
		SCRM3 & +12.93\%$^{-}$ & +9.63\%$^{-}$ & +9.53\%$^{-}$ & +25.15\%$^{-}$ & +23.01\%$^{-}$ & +23.15\%$^{-}$ & +18.65\%$^{-}$ & +16.77\%$^{-}$ & +17.11\%$^{-}$ \\
		\hline
		SCEM & \textbf{+26.59\%} & \textbf{+24.56\%} & \textbf{+26.20\%} & \textbf{+37.43\%} & \textbf{+35.16\%} & \textbf{+37.22\%} & \textbf{+48.99\%} & \textbf{+47.00\%} & \textbf{+50.18\%} \\
		\hline
		\hline
		LSCEM & \textbf{+26.59\%} & \textbf{+24.56\%} & \textbf{+26.20\%} & \textbf{+37.43\%} & \textbf{+35.16\%} & \textbf{+37.22\%} & \textbf{+48.99\%} & \textbf{+47.00\%} & \textbf{+50.18\%} \\
		\hline
		\hline
		\multicolumn{10}{c}{Performance of Re-ranking from the 3rd Page} \\
		\hline
		& \multicolumn{3}{c||}{Toys \& Games} & \multicolumn{3}{c||}{Garden \& Outdoor} & \multicolumn{3}{c}{Cell Phones \& Accessories} \\
		\hline
		Model & $MAP$ & $MRR$ & $NDCG@10$ & $MAP$ & $MRR$ & $NDCG@10$ & $MAP$ & $MRR$ & $NDCG@10$ \\
		\hline
		\hline
		PROD & 0.00\%$^{-}$ & 0.00\%$^{-}$ & 0.00\%$^{-}$ & 0.00\%$^{-}$ & 0.00\%$^{-}$ & 0.00\%$^{-}$ & 0.00\%$^{-}$ & 0.00\%$^{-}$ & 0.00\%$^{-}$ \\
		\hline
		RAND & -15.45\%$^{-}$ & -17.97\%$^{-}$ & -18.96\%$^{-}$ & -12.29\%$^{-}$ & -13.71\%$^{-}$ & -13.97\%$^{-}$ & -8.75\%$^{-}$ & -10.05\%$^{-}$ & -9.55\%$^{-}$ \\
		\hline
		POP & -4.37\%$^{-}$ & -5.31\%$^{-}$ & -5.18\%$^{-}$ & 2.09\%$^{-}$ & 1.43\%$^{-}$ & 3.49\%$^{-}$ & -0.78\%$^{-}$ & -1.21\%$^{-}$ & -1.43\%$^{-}$ \\
		\hline
		QL & -14.87\%$^{-}$ & -18.31\%$^{-}$ & -19.20\%$^{-}$ & -9.15\%$^{-}$ & -10.97\%$^{-}$ & -10.37\%$^{-}$ & -4.05\%$^{-}$ & -5.21\%$^{-}$ & -3.62\%$^{-}$ \\
		\hline
		QEM & +12.83\%$^{-}$ & +11.07\%$^{-}$ & +14.13\%$^{-}$ & +15.82\%$^{-}$ & +14.42\%$^{-}$ & +19.32\%$^{-}$ & +28.85\%$^{-}$ & +27.60\%$^{-}$ & +33.92\%$^{-}$ \\
		\hline
		\hline
		LCRM3 & -13.99\%$^{-}$ & -15.82\%$^{-}$ & -17.20\%$^{-}$ & -9.02\%$^{-}$ & -10.73\%$^{-}$ & -10.04\%$^{-}$ & -3.26\%$^{-}$ & -4.48\%$^{-}$ & -2.85\%$^{-}$ \\
		\hline
		LCEM & +12.83\%$^{-}$ & +11.07\%$^{-}$ & +14.13\%$^{-}$ & +15.82\%$^{-}$ & +14.42\%$^{-}$ & +19.32\%$^{-}$ & +28.85\%$^{-}$ & +27.60\%$^{-}$ & +33.92\%$^{-}$ \\
		\hline
		\hline
		SCRM3 & +34.26\%$^{-}$ & +29.27\%$^{-}$ & +32.86\%$^{-}$ & +49.54\%$^{-}$ & +46.60\%$^{-}$ & +51.20\%$^{-}$ & +44.52\%$^{-}$ & +41.16\%$^{-}$ & +46.98\%$^{-}$ \\
		\hline
		SCEM & \textbf{+51.46\%} & \textbf{+47.57\%} & \textbf{+54.77\%} & \textbf{+63.79\%} & \textbf{+60.43\%} & \textbf{+67.79\%} & \textbf{+85.51\%} & \textbf{+81.72\%} & \textbf{+93.85\%} \\
		\hline
		\hline
		LSCEM & \textbf{+51.46\%} & \textbf{+47.57\%} & \textbf{+54.77\%} & \textbf{+63.79\%} & \textbf{+60.43\%} & \textbf{+67.79\%} & \textbf{+85.51\%} & \textbf{+81.72\%} & \textbf{+93.85\%} \\
		\hline
		\hline
	\end{tabular}
	}
\end{table*}
\footnotetext{Due the confidentiality policy, the absolute value of each metric can not be revealed.}

\subsection{Overall Retrieval Performance}
\label{subsec:overall_perf}
Table \ref{tab:overallperf} shows the performance of different methods on re-ranking items when users paginate to the second and third SERP for \textit{Toys \& Games, Garden \& Outdoor} and \textit{Cell Phones \& Accessories}. Among all the methods, SCEM and SCRM3 perform better than all the other baselines without using short-term context, including their corresponding retrieval baseline, QEM, and QL respectively, and PROD which considers many additional features, showing the effectiveness of incorporating short-term context. 

In contrast to the effectiveness of short-term context, long-term context does not help much when combined with queries alone or together with short-term context. LCRM3 outperforms QL on all the datasets by a small margin when users' historical purchases are used to represent their preferences. LCEM and LSCEM always perform worse than QEM and SCEM by incorporating long-term context with $\lambda_u > 0$. 
Note that since only a small portion of users in the test set appear in the training set, the re-ranking performance of most query sessions in the test set will not be affected. We will elaborate on the effect of long-term context in Section \ref{subsec:user_effect}. 

We found that neural embedding methods are more effective than word-based baselines. When implicit feedback is not incorporated, QEM performs significantly better than QL, sometimes even better than PROD. When clicks are used as context, with neural embeddings, SCEM is much more effective than SCRM3. This shows that semantic match is more beneficial than exact word match for top retrieved items in product search. In addition, these embeddings also carry the popularity information since items purchased more in the training data will get more gradients during training. Due to our model structure, there are also properties that the embeddings of items purchased under similar queries or context will be more alike compared with non-purchased items, and embeddings of clicked and purchased items are also similar. 

The relative improvement of SCEM and SCRM3 compared to the production model on Toys \& Games is less than the other two datasets. There are two possible reasons. 
First, the production model performs better on Toys \& Games, compared with Garden \& Outdoor, and Cell Phones \& Accessories, which can be seen from the larger advantages compared with random re-ranking. 
Second, 
the average clicks in the first two and three SERPs in Toys \& Games are less than the other two datasets \footnote{The specific number of average clicks in the datasets cannot be revealed due to the confidentiality policy. }, thus SCEM and SCRM3 can perform better with more implicit feedback information. 

\begin{figure*}
	\centering
	\begin{subfigure}{.25\textwidth}
		\centering
		\includegraphics[width=1.7in]{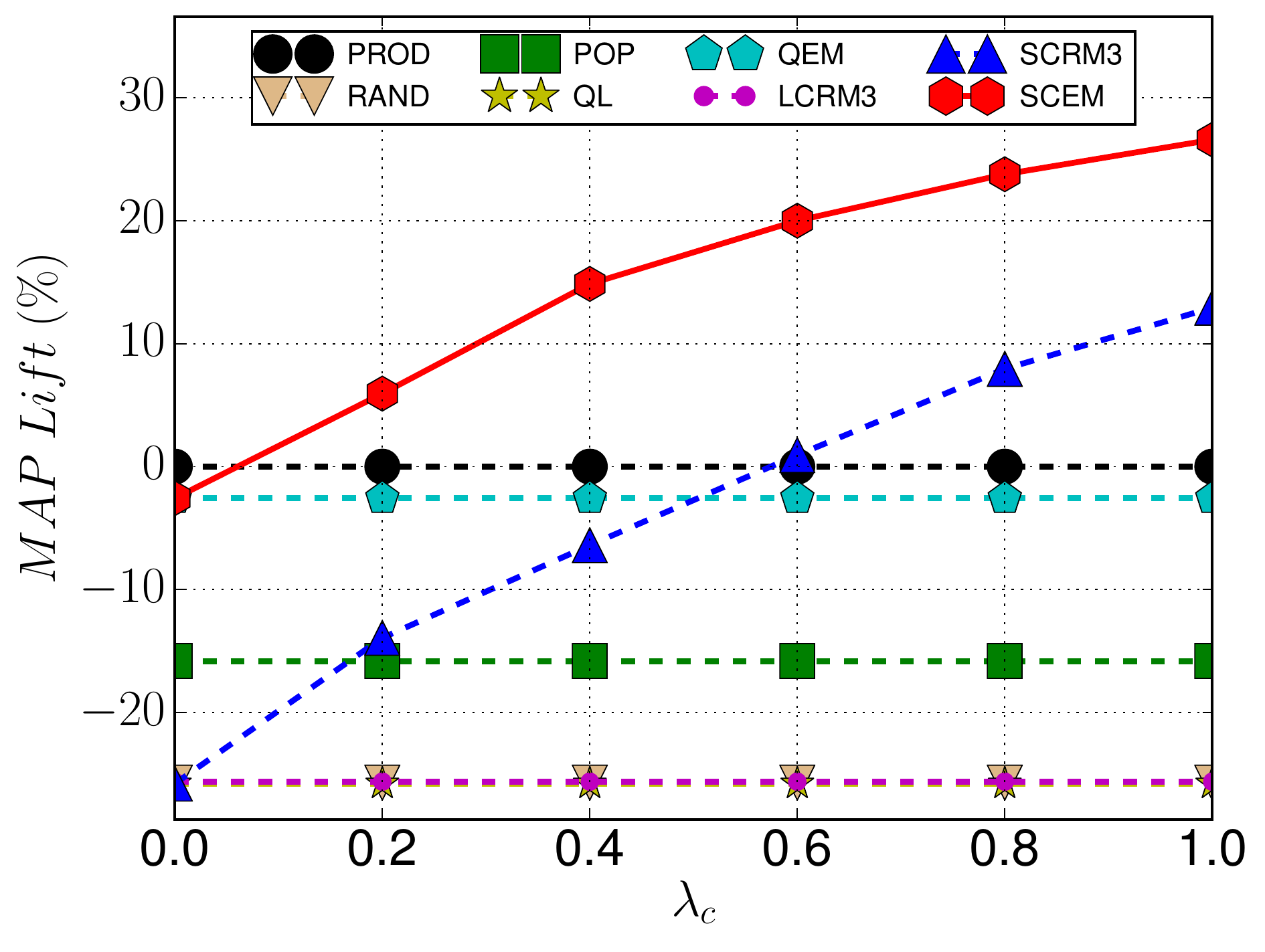}
		\caption{$\lambda_c$}
		\label{fig:a}
	\end{subfigure}%
	\begin{subfigure}{.25\textwidth}
		\centering
		\includegraphics[width=1.7in]{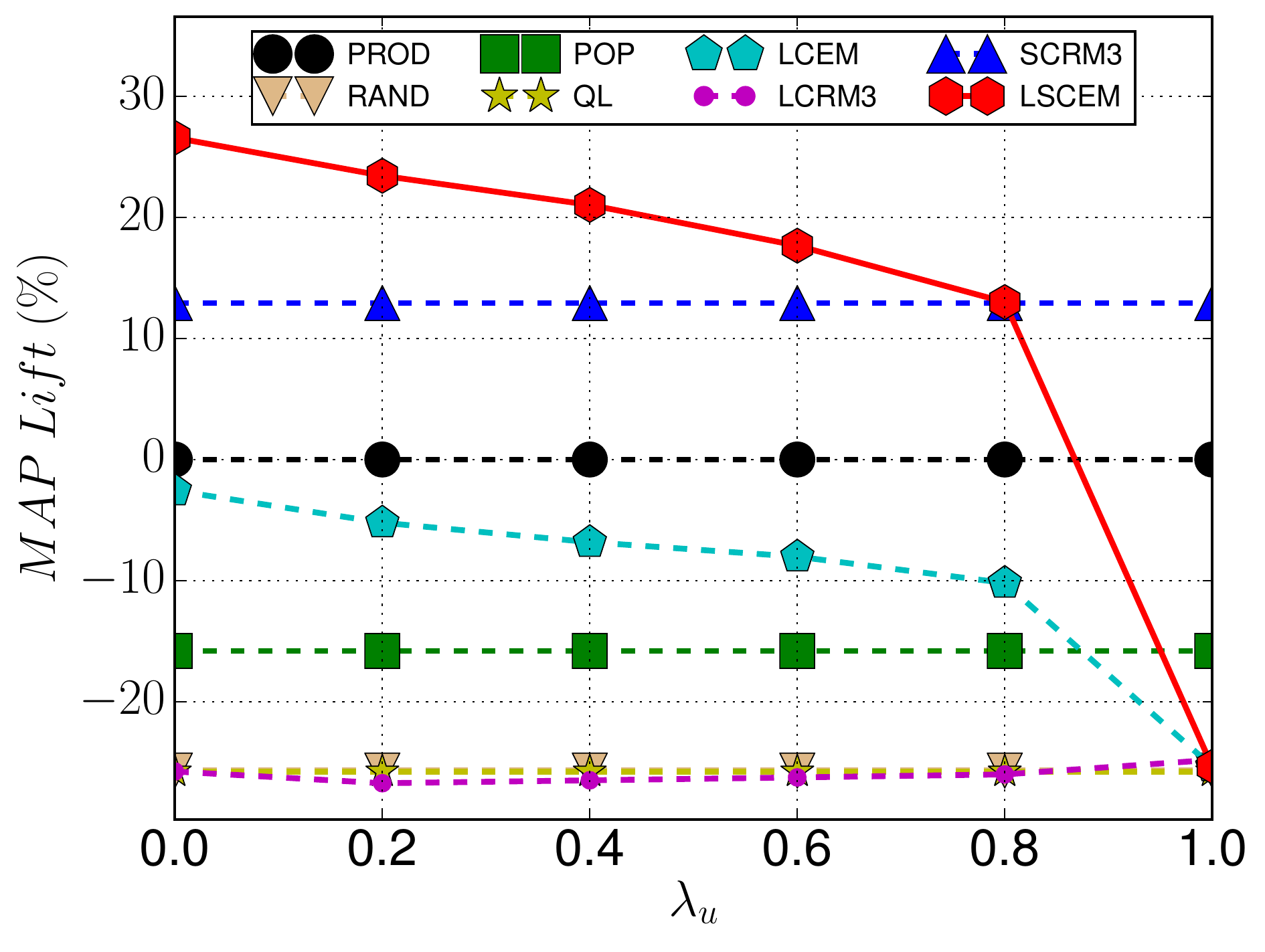}
		\caption{$\lambda_u$}
		\label{fig:b}
	\end{subfigure}%
	\begin{subfigure}{.25\textwidth}
		\centering
		\includegraphics[width=1.7in]{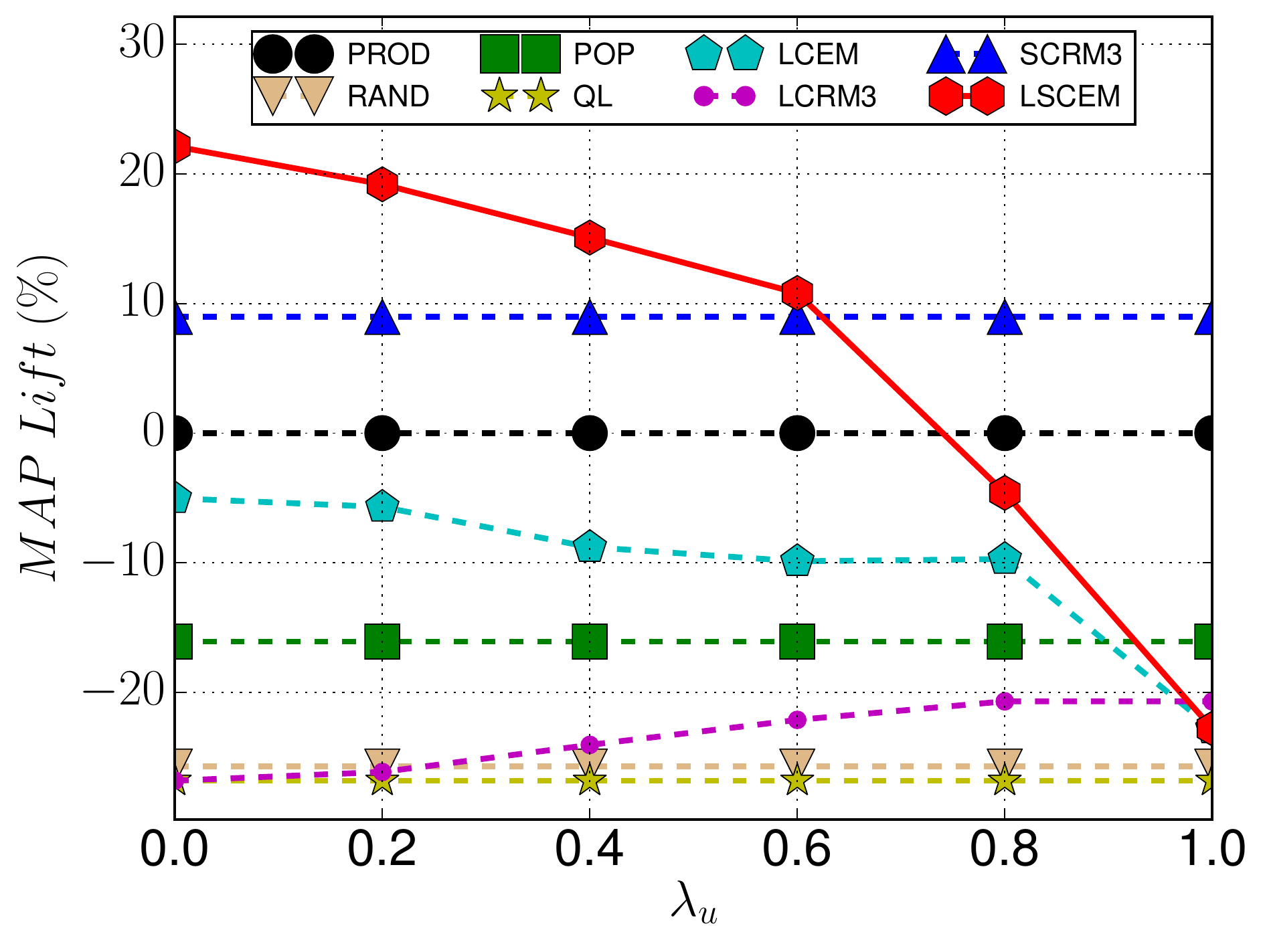}
		\caption{$\lambda_u$ on seen users}
		\label{fig:c}
	\end{subfigure}%
	\begin{subfigure}{.25\textwidth}
		\centering
		\includegraphics[width=1.7in]{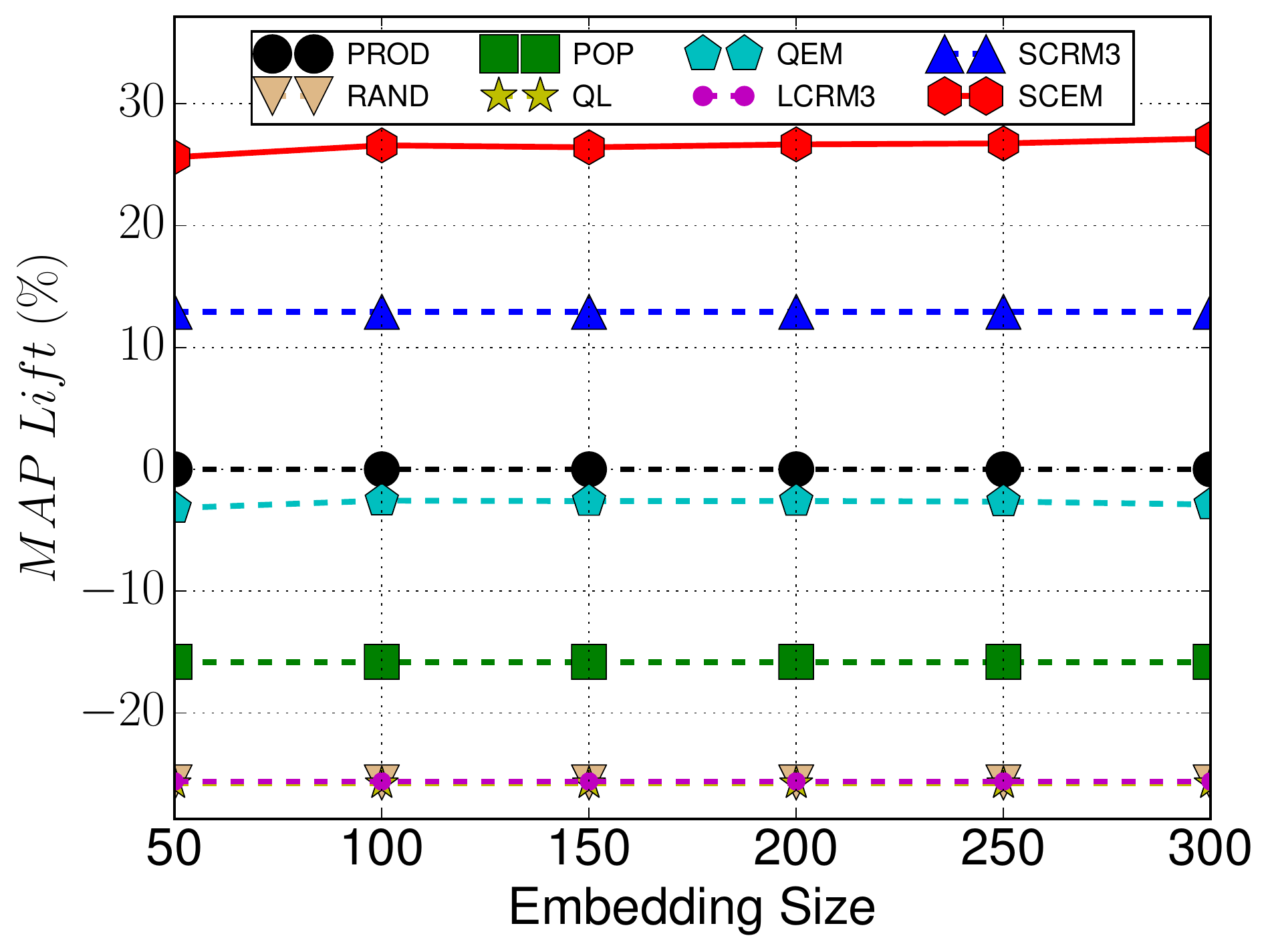}
		\caption{Embedding size}
		\label{fig:d}
	\end{subfigure}%
	\caption{The effect of $\lambda_c$, $\lambda_u$, embedding size on the performance of each model in the collection of \textit{Toys \& Games} when re-ranking from the second SERP for the scenarios where users paginate to page 2.}
	\label{fig:param_effect}
\end{figure*}

The relative performance of all the other methods against PROD is better when re-ranking from page 2 compared with re-ranking from page 3 in terms of all three metrics. Several reasons are shown as follows. 
When purchases happen in the third page or later, it usually means users cannot find the ``right'' products in the first two pages, which further indicates the production model is worse for these query sessions. In addition, the ranking quality of PROD on the third page is worse than on the second page.  
Another reason that SCRM3 and SCEM improve more upon PROD when re-ranking from page 3 is that more context becomes available with clicks collected in the second page and makes the user preference model more robust. 

QL performs similarly to RAND on Toys \& Games and a little better than RAND on Garden \& Outdoor, and Cell Phones \& Accessories, which indicates that relevance captured by exact word matching is not the key concern in the rank lists of the production model.
In addition, most candidate products are consistent with the query intent but the final purchase depends on users' preference. Popularity, as an important factor that consumers will consider, can improve the performance upon QL. However, it is still worse than the production model most of the time.

\subsection{Effect of Short-term Context} 
\label{subsec:query_effect}
We investigate the influence of short-term context by varying the value of $\lambda_c$ with $\lambda_u$ set to 0. The performance of SCRM3 and SCEM varies as the interpolation coefficient of short-term context changes since only these two methods utilize the clicks.
 Since re-ranking from the second or third pages on \textit{Toys \& Games, Garden} and \textit{Mobile} all show similar trends, we only report performance of each method in the setting of re-ranking from second pages on \textit{Toys \& Games}, which is shown in Figure \ref{fig:a}.  
Figure \ref{fig:a} shows that as the weight of clicks is set larger, the performance of SCRM3 and SCEM goes up consistently. When $\lambda_c$ is set to 0, SCRM3 and SCEM degenerate to QL and QEM respectively which do not incorporate short-term context. From another perspective, 
SCRM3 and SCEM degrade in performance as we increase the weight on queries. 
For exact word match based methods, more click signals lead to more improvements for SCRM3, which is also consistent with the fact that QL performs similarly to RAND by only considering queries.
For embedding-based methods which capture semantic match and popularity, QEM with queries alone performs similarly to PROD but much better when more context information is incorporated in SCEM. This indicates that users' clicks already cover the query intent, and also contain additional users' preference information. 

\subsection{Effect of Long-term Context} 
\label{subsec:user_effect}
Next we study the effect of long-term context indicated by users' global representations $\mathcal{E}(u)$ both with and without incorporating short-term context. QEM and LCRM3 only use queries and user historical transactions for ranking; LSCEM uses long and short-term context ($\lambda_u+\lambda_c$ is fixed as 1 since we found that query embeddings do not contribute to the re-ranking performance when short-term context is incorporated).
\textit{Toys \& Games} is used again to show the sensitivity of each model in terms of $\lambda_u$ under the setting of re-ranking from the second page. Since there are users in the test set which never appear in the training set, $\lambda_u$ does not take effect due to the null representations for these unknown users. 
In Toys \& Games, only about 13\% of all the query sessions in the test set are from users who also appear in the training set.
The performance change on the entire test set will be smaller due to the low proportion the models can effect in the test set, so we also include the performance of each model on the subset of data entries associated with users seen in the training set. Figure \ref{fig:b} and \ref{fig:c} show how each method performs on the whole test set and the subset respectively with different $\lambda_u$. 

Figure \ref{fig:b} and \ref{fig:c} show that for LSCEM, as $\lambda_u$ becomes larger,
performance goes down. This indicates that when short-term contexts are used, users' embeddings act like noise and drag down the re-ranking performance. 
$\lambda_u$ has different impacts on the models not using clicks. For LCRM3, when we zoom in to only focus on users that appear in the training set, the performance changes and the superiority over QL are more noticeable. The best value of $MAP$ is achieved when $\lambda_u=0.8$, which means long-term context benefit word-based models with additional information, which can be helpful for solving the word mismatch problem. In contrast, for LCEM, with non-zero $\lambda_u$, it performs worse than only considering queries. Embedding models already capture semantic similarities between words. In addition, as we mentioned in Section \ref{subsec:overall_perf}, they also carry information about popularity since the products purchased more often under the query will get more credits during training. Another possible reason is that the number of customers with sessions of similar intent is low so that the user embedding is misguiding the query sessions.
Thus, users' long-term interests do not bring additional information to further improve LCEM on the collections. 

This finding is different from the observation in HEM proposed by \citet{ai2017learning}, which incorporates user embeddings as users' long-term preferences and achieves superior performance compared to not using user embeddings. 
We hypothesize that this inconsistent finding is due to the differences in datasets.
HEM was experimented on a dataset that is heavily biased to users with multiple purchases and under a rather simplistic assumption of query generation, where the terms from the category hierarchy of a product are concatenated as the query string. Their datasets contain only hundreds of unique queries and tens of thousands items that are all purchased by multiple users. 
In contrast, we experimented on the real queries and corresponding user behavior data extracted from search log. The number of unique queries and items in our experiments are hundreds times larger than in their dataset. There is also little overlap of users in the training and test set in our datasets, while in their experiments, all the users in the test set are shown in the training set. 
 


\subsection{Effect of Embedding Size}
\label{subsec:embed_size}
Figure \ref{fig:d} shows the sensitivity of each model in terms of embedding size on \textit{Toys \& Games}, which presents similar trends to the other two datasets. Generally, SCEM and QEM are not sensitive to the embedding size as long as it is in a reasonable range. 
To keep the model effective and simple, we use 100 as the embedding size and report experimental results under this setting in Table \ref{tab:overallperf} and the other figures. 


\section{Conclusion and Future Work}
\label{sec:conclusion}
We reformulate product search as a dynamic ranking problem where leverage users' implicit feedback on the presented products as short-term context and refine the ranking of remaining items when the users request the next result pages. 
We then propose an end-to-end context-aware neural embedding model to represent various context dependency assumptions for predicting purchased items. Our experimental results indicate that incorporating short-term context is more effective than using long-term context or not using context at all. It is also shown that our neural context-aware model performs better than the state-of-art word-based feedback models. 

For future work, there are several research directions. 
First, it would be better to evaluate our short-term context re-ranking model online, in an interactive setting as each result page can be re-ranked dynamically.
Second, other information sources such as images and price can be included to extract user preferences from their feedback. Third, we are interested in the use of negative feedback such as ``skips'' that can be identified reliably based on subsequent user actions.

\begin{acks}
This work was supported in part by the Center for Intelligent Information Retrieval and in part by NSF IIS-1715095. Any opinions, findings and conclusions or recommendations expressed in this material are those of the authors and do not necessarily reflect those of the sponsor.
\end{acks}

\balance

\bibliographystyle{ACM-Reference-Format}
\bibliography{reference}


\begin{thebibliography}{46}


\ifx \showCODEN    \undefined \def \showCODEN     #1{\unskip}     \fi
\ifx \showDOI      \undefined \def \showDOI       #1{#1}\fi
\ifx \showISBNx    \undefined \def \showISBNx     #1{\unskip}     \fi
\ifx \showISBNxiii \undefined \def \showISBNxiii  #1{\unskip}     \fi
\ifx \showISSN     \undefined \def \showISSN      #1{\unskip}     \fi
\ifx \showLCCN     \undefined \def \showLCCN      #1{\unskip}     \fi
\ifx \shownote     \undefined \def \shownote      #1{#1}          \fi
\ifx \showarticletitle \undefined \def \showarticletitle #1{#1}   \fi
\ifx \showURL      \undefined \def \showURL       {\relax}        \fi
\providecommand\bibfield[2]{#2}
\providecommand\bibinfo[2]{#2}
\providecommand\natexlab[1]{#1}
\providecommand\showeprint[2][]{arXiv:#2}

\bibitem[\protect\citeauthoryear{Ai, Bi, Guo, and Croft}{Ai
  et~al\mbox{.}}{2018a}]%
        {ai2018learning}
\bibfield{author}{\bibinfo{person}{Qingyao Ai}, \bibinfo{person}{Keping Bi},
  \bibinfo{person}{Jiafeng Guo}, {and} \bibinfo{person}{W~Bruce Croft}.}
  \bibinfo{year}{2018}\natexlab{a}.
\newblock \showarticletitle{Learning a Deep Listwise Context Model for Ranking
  Refinement}.
\newblock \bibinfo{journal}{\emph{arXiv preprint arXiv:1804.05936}}
  (\bibinfo{year}{2018}), \bibinfo{pages}{135--144}.
\newblock


\bibitem[\protect\citeauthoryear{Ai, Bi, Luo, Guo, and Croft}{Ai
  et~al\mbox{.}}{2018b}]%
        {ai2018unbiased}
\bibfield{author}{\bibinfo{person}{Qingyao Ai}, \bibinfo{person}{Keping Bi},
  \bibinfo{person}{Cheng Luo}, \bibinfo{person}{Jiafeng Guo}, {and}
  \bibinfo{person}{W~Bruce Croft}.} \bibinfo{year}{2018}\natexlab{b}.
\newblock \showarticletitle{Unbiased Learning to Rank with Unbiased Propensity
  Estimation}.
\newblock \bibinfo{journal}{\emph{arXiv preprint arXiv:1804.05938}}
  (\bibinfo{year}{2018}).
\newblock


\bibitem[\protect\citeauthoryear{Ai, Zhang, Bi, Chen, and Croft}{Ai
  et~al\mbox{.}}{2017}]%
        {ai2017learning}
\bibfield{author}{\bibinfo{person}{Qingyao Ai}, \bibinfo{person}{Yongfeng
  Zhang}, \bibinfo{person}{Keping Bi}, \bibinfo{person}{Xu Chen}, {and}
  \bibinfo{person}{W~Bruce Croft}.} \bibinfo{year}{2017}\natexlab{}.
\newblock \showarticletitle{Learning a hierarchical embedding model for
  personalized product search}. In \bibinfo{booktitle}{\emph{Proceedings of the
  40th International ACM SIGIR Conference}}. ACM, \bibinfo{pages}{645--654}.
\newblock


\bibitem[\protect\citeauthoryear{Chapelle and Zhang}{Chapelle and
  Zhang}{2009}]%
        {chapelle2009dynamic}
\bibfield{author}{\bibinfo{person}{Olivier Chapelle} {and} \bibinfo{person}{Ya
  Zhang}.} \bibinfo{year}{2009}\natexlab{}.
\newblock \showarticletitle{A dynamic bayesian network click model for web
  search ranking}. In \bibinfo{booktitle}{\emph{Proceedings of the 18th
  international conference on World wide web}}. ACM, \bibinfo{pages}{1--10}.
\newblock


\bibitem[\protect\citeauthoryear{Craswell, Zoeter, Taylor, and Ramsey}{Craswell
  et~al\mbox{.}}{2008}]%
        {craswell2008experimental}
\bibfield{author}{\bibinfo{person}{Nick Craswell}, \bibinfo{person}{Onno
  Zoeter}, \bibinfo{person}{Michael Taylor}, {and} \bibinfo{person}{Bill
  Ramsey}.} \bibinfo{year}{2008}\natexlab{}.
\newblock \showarticletitle{An experimental comparison of click position-bias
  models}. In \bibinfo{booktitle}{\emph{Proceedings of the 2008 WSDM
  Conference}}. ACM, \bibinfo{pages}{87--94}.
\newblock


\bibitem[\protect\citeauthoryear{Di, Bhardwaj, Jagadeesh, Piramuthu, and
  Churchill}{Di et~al\mbox{.}}{2014}]%
        {di2014relevance}
\bibfield{author}{\bibinfo{person}{Wei Di}, \bibinfo{person}{Anurag Bhardwaj},
  \bibinfo{person}{Vignesh Jagadeesh}, \bibinfo{person}{Robinson Piramuthu},
  {and} \bibinfo{person}{Elizabeth Churchill}.}
  \bibinfo{year}{2014}\natexlab{}.
\newblock \showarticletitle{When relevance is not enough: Promoting visual
  attractiveness for fashion e-commerce}.
\newblock \bibinfo{journal}{\emph{arXiv preprint arXiv:1406.3561}}
  (\bibinfo{year}{2014}).
\newblock


\bibitem[\protect\citeauthoryear{Duan, Zhai, Cheng, and Gattani}{Duan
  et~al\mbox{.}}{2013a}]%
        {duan2013probabilistic}
\bibfield{author}{\bibinfo{person}{Huizhong Duan}, \bibinfo{person}{ChengXiang
  Zhai}, \bibinfo{person}{Jinxing Cheng}, {and} \bibinfo{person}{Abhishek
  Gattani}.} \bibinfo{year}{2013}\natexlab{a}.
\newblock \showarticletitle{A probabilistic mixture model for mining and
  analyzing product search log}. In \bibinfo{booktitle}{\emph{Proceedings of
  the 22nd ACM international conference on Information \& Knowledge
  Management}}. ACM, \bibinfo{pages}{2179--2188}.
\newblock


\bibitem[\protect\citeauthoryear{Duan, Zhai, Cheng, and Gattani}{Duan
  et~al\mbox{.}}{2013b}]%
        {duan2013supporting}
\bibfield{author}{\bibinfo{person}{Huizhong Duan}, \bibinfo{person}{ChengXiang
  Zhai}, \bibinfo{person}{Jinxing Cheng}, {and} \bibinfo{person}{Abhishek
  Gattani}.} \bibinfo{year}{2013}\natexlab{b}.
\newblock \showarticletitle{Supporting keyword search in product database: a
  probabilistic approach}.
\newblock \bibinfo{journal}{\emph{Proceedings of the VLDB Endowment}}
  \bibinfo{volume}{6}, \bibinfo{number}{14} (\bibinfo{year}{2013}),
  \bibinfo{pages}{1786--1797}.
\newblock


\bibitem[\protect\citeauthoryear{Dupret and Piwowarski}{Dupret and
  Piwowarski}{2008}]%
        {dupret2008user}
\bibfield{author}{\bibinfo{person}{Georges~E Dupret} {and}
  \bibinfo{person}{Benjamin Piwowarski}.} \bibinfo{year}{2008}\natexlab{}.
\newblock \showarticletitle{A user browsing model to predict search engine
  click data from past observations.}. In \bibinfo{booktitle}{\emph{Proceedings
  of the 31st annual international ACM SIGIR conference}}. ACM,
  \bibinfo{pages}{331--338}.
\newblock


\bibitem[\protect\citeauthoryear{Garcia}{Garcia}{2018}]%
        {Garcia2018}
\bibfield{author}{\bibinfo{person}{Krista Garcia}.}
  \bibinfo{year}{2018}\natexlab{}.
\newblock \bibinfo{title}{More Product Searches Start on Amazon}.
\newblock
\newblock
\urldef\tempurl%
\url{https://retail.emarketer.com/article/more-product-searches-start-on-amazon/5b92c0e0ebd40005bc4dc7ae}
\showURL{%
\tempurl}


\bibitem[\protect\citeauthoryear{Guo, Cheng, Nie, Xu, and Kankanhalli}{Guo
  et~al\mbox{.}}{2018}]%
        {guo2018multi}
\bibfield{author}{\bibinfo{person}{Yangyang Guo}, \bibinfo{person}{Zhiyong
  Cheng}, \bibinfo{person}{Liqiang Nie}, \bibinfo{person}{Xin-Shun Xu}, {and}
  \bibinfo{person}{Mohan Kankanhalli}.} \bibinfo{year}{2018}\natexlab{}.
\newblock \showarticletitle{Multi-modal preference modeling for product
  search}. In \bibinfo{booktitle}{\emph{2018 ACM Multimedia Conference on
  Multimedia Conference}}. ACM, \bibinfo{pages}{1865--1873}.
\newblock


\bibitem[\protect\citeauthoryear{Hidasi, Karatzoglou, Baltrunas, and
  Tikk}{Hidasi et~al\mbox{.}}{2015}]%
        {hidasi2015session}
\bibfield{author}{\bibinfo{person}{Bal{\'a}zs Hidasi},
  \bibinfo{person}{Alexandros Karatzoglou}, \bibinfo{person}{Linas Baltrunas},
  {and} \bibinfo{person}{Domonkos Tikk}.} \bibinfo{year}{2015}\natexlab{}.
\newblock \showarticletitle{Session-based recommendations with recurrent neural
  networks}.
\newblock \bibinfo{journal}{\emph{arXiv preprint arXiv:1511.06939}}
  (\bibinfo{year}{2015}).
\newblock


\bibitem[\protect\citeauthoryear{Hidasi and Tikk}{Hidasi and Tikk}{2016}]%
        {hidasi2016general}
\bibfield{author}{\bibinfo{person}{Bal{\'a}zs Hidasi} {and}
  \bibinfo{person}{Domonkos Tikk}.} \bibinfo{year}{2016}\natexlab{}.
\newblock \showarticletitle{General factorization framework for context-aware
  recommendations}.
\newblock \bibinfo{journal}{\emph{Data Mining and Knowledge Discovery}}
  \bibinfo{volume}{30}, \bibinfo{number}{2} (\bibinfo{year}{2016}),
  \bibinfo{pages}{342--371}.
\newblock


\bibitem[\protect\citeauthoryear{Hu, Da, Zeng, Yu, and Xu}{Hu
  et~al\mbox{.}}{2018}]%
        {hu2018reinforcement}
\bibfield{author}{\bibinfo{person}{Yujing Hu}, \bibinfo{person}{Qing Da},
  \bibinfo{person}{Anxiang Zeng}, \bibinfo{person}{Yang Yu}, {and}
  \bibinfo{person}{Yinghui Xu}.} \bibinfo{year}{2018}\natexlab{}.
\newblock \showarticletitle{Reinforcement Learning to Rank in E-Commerce Search
  Engine: Formalization, Analysis, and Application}.
\newblock \bibinfo{journal}{\emph{arXiv preprint arXiv:1803.00710}}
  (\bibinfo{year}{2018}).
\newblock


\bibitem[\protect\citeauthoryear{Jansen and Molina}{Jansen and Molina}{2006}]%
        {jansen2006effectiveness}
\bibfield{author}{\bibinfo{person}{Bernard~J Jansen} {and}
  \bibinfo{person}{Paulo~R Molina}.} \bibinfo{year}{2006}\natexlab{}.
\newblock \showarticletitle{The effectiveness of Web search engines for
  retrieving relevant ecommerce links}.
\newblock \bibinfo{journal}{\emph{Information Processing \& Management}}
  \bibinfo{volume}{42}, \bibinfo{number}{4} (\bibinfo{year}{2006}),
  \bibinfo{pages}{1075--1098}.
\newblock


\bibitem[\protect\citeauthoryear{Jawaheer, Weller, and Kostkova}{Jawaheer
  et~al\mbox{.}}{2014}]%
        {jawaheer2014modeling}
\bibfield{author}{\bibinfo{person}{Gawesh Jawaheer}, \bibinfo{person}{Peter
  Weller}, {and} \bibinfo{person}{Patty Kostkova}.}
  \bibinfo{year}{2014}\natexlab{}.
\newblock \showarticletitle{Modeling user preferences in recommender systems: A
  classification framework for explicit and implicit user feedback}.
\newblock \bibinfo{journal}{\emph{ACM Transactions on Interactive Intelligent
  Systems (TiiS)}} \bibinfo{volume}{4}, \bibinfo{number}{2}
  (\bibinfo{year}{2014}), \bibinfo{pages}{8}.
\newblock


\bibitem[\protect\citeauthoryear{Joachims, Granka, Pan, Hembrooke, and
  Gay}{Joachims et~al\mbox{.}}{2017a}]%
        {joachims2017accurately}
\bibfield{author}{\bibinfo{person}{Thorsten Joachims}, \bibinfo{person}{Laura
  Granka}, \bibinfo{person}{Bing Pan}, \bibinfo{person}{Helene Hembrooke},
  {and} \bibinfo{person}{Geri Gay}.} \bibinfo{year}{2017}\natexlab{a}.
\newblock \showarticletitle{Accurately interpreting clickthrough data as
  implicit feedback}. In \bibinfo{booktitle}{\emph{ACM SIGIR Forum}},
  Vol.~\bibinfo{volume}{51}. Acm, \bibinfo{pages}{4--11}.
\newblock


\bibitem[\protect\citeauthoryear{Joachims, Swaminathan, and Schnabel}{Joachims
  et~al\mbox{.}}{2017b}]%
        {joachims2017unbiased}
\bibfield{author}{\bibinfo{person}{Thorsten Joachims}, \bibinfo{person}{Adith
  Swaminathan}, {and} \bibinfo{person}{Tobias Schnabel}.}
  \bibinfo{year}{2017}\natexlab{b}.
\newblock \showarticletitle{Unbiased learning-to-rank with biased feedback}. In
  \bibinfo{booktitle}{\emph{Proceedings of the Tenth ACM International
  Conference on Web Search and Data Mining}}. ACM, \bibinfo{pages}{781--789}.
\newblock


\bibitem[\protect\citeauthoryear{Karmaker~Santu, Sondhi, and
  Zhai}{Karmaker~Santu et~al\mbox{.}}{2017}]%
        {karmaker2017application}
\bibfield{author}{\bibinfo{person}{Shubhra~Kanti Karmaker~Santu},
  \bibinfo{person}{Parikshit Sondhi}, {and} \bibinfo{person}{ChengXiang Zhai}.}
  \bibinfo{year}{2017}\natexlab{}.
\newblock \showarticletitle{On application of learning to rank for e-commerce
  search}. In \bibinfo{booktitle}{\emph{Proceedings of the 40th International
  ACM SIGIR Conference}}. ACM, \bibinfo{pages}{475--484}.
\newblock


\bibitem[\protect\citeauthoryear{Kingma and Ba}{Kingma and Ba}{2014}]%
        {kingma2014adam}
\bibfield{author}{\bibinfo{person}{Diederik~P Kingma} {and}
  \bibinfo{person}{Jimmy~Lei Ba}.} \bibinfo{year}{2014}\natexlab{}.
\newblock \showarticletitle{Adam: Amethod for stochastic optimization}. In
  \bibinfo{booktitle}{\emph{Proc. 3rd Int. Conf. Learn. Representations}}.
\newblock


\bibitem[\protect\citeauthoryear{Lavrenko and Croft}{Lavrenko and
  Croft}{2017}]%
        {lavrenko2017relevance}
\bibfield{author}{\bibinfo{person}{Victor Lavrenko} {and}
  \bibinfo{person}{W~Bruce Croft}.} \bibinfo{year}{2017}\natexlab{}.
\newblock \showarticletitle{Relevance-based language models}. In
  \bibinfo{booktitle}{\emph{ACM SIGIR Forum}}, Vol.~\bibinfo{volume}{51}. ACM,
  \bibinfo{pages}{260--267}.
\newblock


\bibitem[\protect\citeauthoryear{Li, Ghose, and Ipeirotis}{Li
  et~al\mbox{.}}{2011}]%
        {li2011towards}
\bibfield{author}{\bibinfo{person}{Beibei Li}, \bibinfo{person}{Anindya Ghose},
  {and} \bibinfo{person}{Panagiotis~G Ipeirotis}.}
  \bibinfo{year}{2011}\natexlab{}.
\newblock \showarticletitle{Towards a theory model for product search}. In
  \bibinfo{booktitle}{\emph{Proceedings of the 20th international conference on
  World wide web}}. ACM, \bibinfo{pages}{327--336}.
\newblock


\bibitem[\protect\citeauthoryear{Li, Ren, Chen, Ren, Lian, and Ma}{Li
  et~al\mbox{.}}{2017}]%
        {li2017neural}
\bibfield{author}{\bibinfo{person}{Jing Li}, \bibinfo{person}{Pengjie Ren},
  \bibinfo{person}{Zhumin Chen}, \bibinfo{person}{Zhaochun Ren},
  \bibinfo{person}{Tao Lian}, {and} \bibinfo{person}{Jun Ma}.}
  \bibinfo{year}{2017}\natexlab{}.
\newblock \showarticletitle{Neural attentive session-based recommendation}. In
  \bibinfo{booktitle}{\emph{Proceedings of the 2017 ACM on Conference on
  Information and Knowledge Management}}. ACM, \bibinfo{pages}{1419--1428}.
\newblock


\bibitem[\protect\citeauthoryear{Lim, Liu, and Lee}{Lim et~al\mbox{.}}{2010}]%
        {lim2010multi}
\bibfield{author}{\bibinfo{person}{Soon Chong~Johnson Lim},
  \bibinfo{person}{Ying Liu}, {and} \bibinfo{person}{Wing~Bun Lee}.}
  \bibinfo{year}{2010}\natexlab{}.
\newblock \showarticletitle{Multi-facet product information search and
  retrieval using semantically annotated product family ontology}.
\newblock \bibinfo{journal}{\emph{Information Processing \& Management}}
  \bibinfo{volume}{46}, \bibinfo{number}{4} (\bibinfo{year}{2010}),
  \bibinfo{pages}{479--493}.
\newblock


\bibitem[\protect\citeauthoryear{Long, Bian, Dong, and Chang}{Long
  et~al\mbox{.}}{2012}]%
        {long2012enhancing}
\bibfield{author}{\bibinfo{person}{Bo Long}, \bibinfo{person}{Jiang Bian},
  \bibinfo{person}{Anlei Dong}, {and} \bibinfo{person}{Yi Chang}.}
  \bibinfo{year}{2012}\natexlab{}.
\newblock \showarticletitle{Enhancing product search by best-selling prediction
  in e-commerce}. In \bibinfo{booktitle}{\emph{Proceedings of the 21st ACM CIKM
  Conference}}. ACM, \bibinfo{pages}{2479--2482}.
\newblock


\bibitem[\protect\citeauthoryear{Lv and Zhai}{Lv and Zhai}{2009}]%
        {lv2009adaptive}
\bibfield{author}{\bibinfo{person}{Yuanhua Lv} {and}
  \bibinfo{person}{ChengXiang Zhai}.} \bibinfo{year}{2009}\natexlab{}.
\newblock \showarticletitle{Adaptive relevance feedback in information
  retrieval}. In \bibinfo{booktitle}{\emph{Proceedings of the 18th ACM
  conference on Information and knowledge management}}. ACM,
  \bibinfo{pages}{255--264}.
\newblock


\bibitem[\protect\citeauthoryear{Parikh and Sundaresan}{Parikh and
  Sundaresan}{2011}]%
        {parikh2011beyond}
\bibfield{author}{\bibinfo{person}{Nish Parikh} {and} \bibinfo{person}{Neel
  Sundaresan}.} \bibinfo{year}{2011}\natexlab{}.
\newblock \showarticletitle{Beyond relevance in marketplace search}. In
  \bibinfo{booktitle}{\emph{Proceedings of the 20th ACM CIKM Conference}}. ACM,
  \bibinfo{pages}{2109--2112}.
\newblock


\bibitem[\protect\citeauthoryear{Ponte and Croft}{Ponte and Croft}{1998}]%
        {ponte1998language}
\bibfield{author}{\bibinfo{person}{Jay~M Ponte} {and} \bibinfo{person}{W~Bruce
  Croft}.} \bibinfo{year}{1998}\natexlab{}.
\newblock \showarticletitle{A language modeling approach to information
  retrieval}. In \bibinfo{booktitle}{\emph{Proceedings of the 21st annual
  international ACM SIGIR conference}}. ACM, \bibinfo{pages}{275--281}.
\newblock


\bibitem[\protect\citeauthoryear{Quadrana, Cremonesi, and Jannach}{Quadrana
  et~al\mbox{.}}{2018}]%
        {quadrana2018sequence}
\bibfield{author}{\bibinfo{person}{Massimo Quadrana}, \bibinfo{person}{Paolo
  Cremonesi}, {and} \bibinfo{person}{Dietmar Jannach}.}
  \bibinfo{year}{2018}\natexlab{}.
\newblock \showarticletitle{Sequence-Aware Recommender Systems}.
\newblock \bibinfo{journal}{\emph{ACM Comput. Surv.}} (\bibinfo{year}{2018}).
\newblock


\bibitem[\protect\citeauthoryear{Quadrana, Karatzoglou, Hidasi, and
  Cremonesi}{Quadrana et~al\mbox{.}}{2017}]%
        {quadrana2017personalizing}
\bibfield{author}{\bibinfo{person}{Massimo Quadrana},
  \bibinfo{person}{Alexandros Karatzoglou}, \bibinfo{person}{Bal{\'a}zs
  Hidasi}, {and} \bibinfo{person}{Paolo Cremonesi}.}
  \bibinfo{year}{2017}\natexlab{}.
\newblock \showarticletitle{Personalizing session-based recommendations with
  hierarchical recurrent neural networks}. In
  \bibinfo{booktitle}{\emph{Proceedings of the Eleventh ACM Conference on
  Recommender Systems}}. ACM, \bibinfo{pages}{130--137}.
\newblock


\bibitem[\protect\citeauthoryear{Rekabsaz, Lupu, Hanbury, and Zuccon}{Rekabsaz
  et~al\mbox{.}}{2016}]%
        {Rekabsaz:2016:GTM:2983323.2983833}
\bibfield{author}{\bibinfo{person}{Navid Rekabsaz}, \bibinfo{person}{Mihai
  Lupu}, \bibinfo{person}{Allan Hanbury}, {and} \bibinfo{person}{Guido
  Zuccon}.} \bibinfo{year}{2016}\natexlab{}.
\newblock \showarticletitle{Generalizing translation models in the
  probabilistic relevance framework}. In \bibinfo{booktitle}{\emph{Proceedings
  of the 25th ACM CIKM conference}}. ACM, \bibinfo{pages}{711--720}.
\newblock


\bibitem[\protect\citeauthoryear{Rendle, Gantner, Freudenthaler, and
  Schmidt-Thieme}{Rendle et~al\mbox{.}}{2011}]%
        {rendle2011fast}
\bibfield{author}{\bibinfo{person}{Steffen Rendle}, \bibinfo{person}{Zeno
  Gantner}, \bibinfo{person}{Christoph Freudenthaler}, {and}
  \bibinfo{person}{Lars Schmidt-Thieme}.} \bibinfo{year}{2011}\natexlab{}.
\newblock \showarticletitle{Fast context-aware recommendations with
  factorization machines}. In \bibinfo{booktitle}{\emph{Proceedings of the 34th
  international ACM SIGIR Conference}}. ACM, \bibinfo{pages}{635--644}.
\newblock


\bibitem[\protect\citeauthoryear{Rocchio}{Rocchio}{1971}]%
        {rocchio1971relevance}
\bibfield{author}{\bibinfo{person}{Joseph~John Rocchio}.}
  \bibinfo{year}{1971}\natexlab{}.
\newblock \showarticletitle{Relevance feedback in information retrieval}.
\newblock \bibinfo{journal}{\emph{The Smart retrieval system-experiments in
  automatic document processing}} (\bibinfo{year}{1971}).
\newblock


\bibitem[\protect\citeauthoryear{Saleh}{Saleh}{2018}]%
        {saleh2018}
\bibfield{author}{\bibinfo{person}{Khalid Saleh}.}
  \bibinfo{year}{2018}\natexlab{}.
\newblock \bibinfo{title}{Global Online Retail Spending - Statistics and
  Trends}.
\newblock
\newblock
\urldef\tempurl%
\url{https://www.invespcro.com/blog/global-online-retail-spending-statistics-and-trends/}
\showURL{%
\tempurl}


\bibitem[\protect\citeauthoryear{Salton, Wong, and Yang}{Salton
  et~al\mbox{.}}{1975}]%
        {salton1975vector}
\bibfield{author}{\bibinfo{person}{Gerard Salton}, \bibinfo{person}{Anita
  Wong}, {and} \bibinfo{person}{Chung-Shu Yang}.}
  \bibinfo{year}{1975}\natexlab{}.
\newblock \showarticletitle{A vector space model for automatic indexing}.
\newblock \bibinfo{journal}{\emph{Commun. ACM}} \bibinfo{volume}{18},
  \bibinfo{number}{11} (\bibinfo{year}{1975}), \bibinfo{pages}{613--620}.
\newblock


\bibitem[\protect\citeauthoryear{Taylor, Guiver, Robertson, and Minka}{Taylor
  et~al\mbox{.}}{2008}]%
        {taylor2008softrank}
\bibfield{author}{\bibinfo{person}{Michael Taylor}, \bibinfo{person}{John
  Guiver}, \bibinfo{person}{Stephen Robertson}, {and} \bibinfo{person}{Tom
  Minka}.} \bibinfo{year}{2008}\natexlab{}.
\newblock \showarticletitle{Softrank: optimizing non-smooth rank metrics}. In
  \bibinfo{booktitle}{\emph{Proceedings of the 2008 International Conference on
  Web Search and Data Mining}}. ACM, \bibinfo{pages}{77--86}.
\newblock


\bibitem[\protect\citeauthoryear{Twardowski}{Twardowski}{2016}]%
        {twardowski2016modelling}
\bibfield{author}{\bibinfo{person}{Bart{\l}omiej Twardowski}.}
  \bibinfo{year}{2016}\natexlab{}.
\newblock \showarticletitle{Modelling contextual information in session-aware
  recommender systems with neural networks}. In
  \bibinfo{booktitle}{\emph{Proceedings of the 10th ACM Conference on
  Recommender Systems}}. ACM, \bibinfo{pages}{273--276}.
\newblock


\bibitem[\protect\citeauthoryear{Van~Gysel, de~Rijke, and Kanoulas}{Van~Gysel
  et~al\mbox{.}}{2016}]%
        {van2016learning}
\bibfield{author}{\bibinfo{person}{Christophe Van~Gysel},
  \bibinfo{person}{Maarten de Rijke}, {and} \bibinfo{person}{Evangelos
  Kanoulas}.} \bibinfo{year}{2016}\natexlab{}.
\newblock \showarticletitle{Learning latent vector spaces for product search}.
  In \bibinfo{booktitle}{\emph{Proceedings of the 25th ACM CIKM Conference}}.
  ACM, \bibinfo{pages}{165--174}.
\newblock


\bibitem[\protect\citeauthoryear{Vandic, Frasincar, and Kaymak}{Vandic
  et~al\mbox{.}}{2013}]%
        {vandic2013facet}
\bibfield{author}{\bibinfo{person}{Damir Vandic}, \bibinfo{person}{Flavius
  Frasincar}, {and} \bibinfo{person}{Uzay Kaymak}.}
  \bibinfo{year}{2013}\natexlab{}.
\newblock \showarticletitle{Facet selection algorithms for web product search}.
  In \bibinfo{booktitle}{\emph{Proceedings of the 22nd ACM CIKM Conference}}.
  ACM, \bibinfo{pages}{2327--2332}.
\newblock


\bibitem[\protect\citeauthoryear{Wang, Golbandi, Bendersky, Metzler, and
  Najork}{Wang et~al\mbox{.}}{2018}]%
        {wang2018position}
\bibfield{author}{\bibinfo{person}{Xuanhui Wang}, \bibinfo{person}{Nadav
  Golbandi}, \bibinfo{person}{Michael Bendersky}, \bibinfo{person}{Donald
  Metzler}, {and} \bibinfo{person}{Marc Najork}.}
  \bibinfo{year}{2018}\natexlab{}.
\newblock \showarticletitle{Position bias estimation for unbiased learning to
  rank in personal search}. In \bibinfo{booktitle}{\emph{Proceedings of the
  Eleventh ACM WSDM Conference}}. ACM, \bibinfo{pages}{610--618}.
\newblock


\bibitem[\protect\citeauthoryear{Wu and Yan}{Wu and Yan}{2017}]%
        {wu2017session}
\bibfield{author}{\bibinfo{person}{Chen Wu} {and} \bibinfo{person}{Ming Yan}.}
  \bibinfo{year}{2017}\natexlab{}.
\newblock \showarticletitle{Session-aware information embedding for e-commerce
  product recommendation}. In \bibinfo{booktitle}{\emph{Proceedings of the 2017
  ACM on Conference on Information and Knowledge Management}}. ACM,
  \bibinfo{pages}{2379--2382}.
\newblock


\bibitem[\protect\citeauthoryear{Wu, Hu, Hong, and Liu}{Wu
  et~al\mbox{.}}{2018}]%
        {wu2018turning}
\bibfield{author}{\bibinfo{person}{Liang Wu}, \bibinfo{person}{Diane Hu},
  \bibinfo{person}{Liangjie Hong}, {and} \bibinfo{person}{Huan Liu}.}
  \bibinfo{year}{2018}\natexlab{}.
\newblock \showarticletitle{Turning Clicks into Purchases: Revenue Optimization
  for Product Search in E-Commerce}.
\newblock  (\bibinfo{year}{2018}).
\newblock


\bibitem[\protect\citeauthoryear{Xia, Liu, Wang, Zhang, and Li}{Xia
  et~al\mbox{.}}{2008}]%
        {xia2008listwise}
\bibfield{author}{\bibinfo{person}{Fen Xia}, \bibinfo{person}{Tie-Yan Liu},
  \bibinfo{person}{Jue Wang}, \bibinfo{person}{Wensheng Zhang}, {and}
  \bibinfo{person}{Hang Li}.} \bibinfo{year}{2008}\natexlab{}.
\newblock \showarticletitle{Listwise approach to learning to rank: theory and
  algorithm}. In \bibinfo{booktitle}{\emph{Proceedings of the 25th
  international conference on Machine learning}}. ACM,
  \bibinfo{pages}{1192--1199}.
\newblock


\bibitem[\protect\citeauthoryear{Yu, Mohan, Putthividhya, and Wong}{Yu
  et~al\mbox{.}}{2014}]%
        {yu2014latent}
\bibfield{author}{\bibinfo{person}{Jun Yu}, \bibinfo{person}{Sunil Mohan},
  \bibinfo{person}{Duangmanee~Pew Putthividhya}, {and}
  \bibinfo{person}{Weng-Keen Wong}.} \bibinfo{year}{2014}\natexlab{}.
\newblock \showarticletitle{Latent dirichlet allocation based diversified
  retrieval for e-commerce search}. In \bibinfo{booktitle}{\emph{Proceedings of
  the 7th ACM WSDM Conference}}. ACM, \bibinfo{pages}{463--472}.
\newblock


\bibitem[\protect\citeauthoryear{Yue and Joachims}{Yue and Joachims}{2009}]%
        {yue2009interactively}
\bibfield{author}{\bibinfo{person}{Yisong Yue} {and} \bibinfo{person}{Thorsten
  Joachims}.} \bibinfo{year}{2009}\natexlab{}.
\newblock \showarticletitle{Interactively optimizing information retrieval
  systems as a dueling bandits problem}. In
  \bibinfo{booktitle}{\emph{Proceedings of the 26th Annual International
  Conference on Machine Learning}}. ACM, \bibinfo{pages}{1201--1208}.
\newblock


\bibitem[\protect\citeauthoryear{Zamani and Croft}{Zamani and Croft}{2016}]%
        {Zamani:2016:EQL:2970398.2970405}
\bibfield{author}{\bibinfo{person}{Hamed Zamani} {and} \bibinfo{person}{W~Bruce
  Croft}.} \bibinfo{year}{2016}\natexlab{}.
\newblock \showarticletitle{Embedding-based query language models}. In
  \bibinfo{booktitle}{\emph{Proceedings of the 2016 ACM ICTIR conference}}.
  ACM, \bibinfo{pages}{147--156}.
\newblock


\end{thebibliography}

\end{document}